\documentclass[11pt]{article}

\usepackage[T1]{fontenc}
\usepackage[utf8]{inputenc}
\usepackage{mathptmx}
\usepackage[a4paper,margin=1in]{geometry}
\usepackage{amsmath,amssymb,mathtools,bm}
\usepackage{graphicx}
\usepackage{booktabs,tabularx,multirow,threeparttable,array}
\usepackage[table]{xcolor}
\usepackage{tikz}
\usetikzlibrary{arrows.meta,positioning,fit,backgrounds,calc,shapes.geometric}
\usepackage{algorithm}
\usepackage[noend]{algpseudocode}
\usepackage[font=small,labelfont=bf,skip=4pt]{caption}
\usepackage{enumitem}
\usepackage[numbers,sort&compress]{natbib}
\usepackage{microtype}
\usepackage{xurl}
\usepackage[colorlinks=true,linkcolor=blue!55!black,citecolor=blue!55!black,urlcolor=blue!55!black]{hyperref}
\usepackage[capitalise,noabbrev]{cleveref}

\usepackage[section]{placeins}
\setlist{itemsep=2pt,topsep=3pt}

\newcolumntype{L}{>{\raggedright\arraybackslash}X}
\newcolumntype{C}{>{\centering\arraybackslash}X}
\crefname{algorithm}{Algorithm}{Algorithms}

\crefname{hypothesis}{Hypothesis}{Hypotheses}

\newcommand{\AGC}{\textup{\textsc{agc}}}
\newcommand{\AGT}{\textup{\textsc{agt}}}
\newcommand{\DF}{\mathrm{DF}}
\newcommand{\CLT}{\mathrm{CLT}}
\newcommand{\CFR}{\mathrm{CFR}}
\newcommand{\FDRT}{\mathrm{FDRT}}
\newcommand{\DRR}{\mathrm{DRR}}
\newcommand{\FI}{\mathrm{FI}}
\newcommand{\E}{\mathbb{E}}
\newcommand{\1}{\mathbb{1}}
\DeclareMathOperator{\median}{median}
\DeclareMathOperator{\LN}{LogNormal}
\DeclareMathOperator{\Pois}{Poisson}
\DeclareMathOperator{\Bern}{Bernoulli}

\DeclareMathOperator{\Expo}{Exp}

\title{\textbf{How AI Changes DevOps Performance: A Mechanism-Based Simulation}}

\author{Mamdouh Alenezi\\[2pt]
\small SDAIA Academy, Saudi Data and Artificial Intelligence Authority (SDAIA), Riyadh, Saudi Arabia}

\date{}

\begin{document}
\maketitle

\begin{abstract}
\noindent\textbf{Context.} AI tools for software development have progressed from code completion to autonomous agents, and organisations judge their value with DORA delivery metrics. The evidence is contradictory, mostly cross-sectional, and treats AI-generated code and agentic AI as one construct.
\textbf{Objective.} We examine how AI-generated code (\AGC{}) and agentic AI (\AGT{}) separately and jointly affect deployment frequency, change lead time, change failure rate, failed deployment recovery time and deployment rework rate.
\textbf{Method.} We audit the quantitative evidence, formalise a two-construct model with four capability moderators, and implement it as a discrete-event simulation of a delivery pipeline. Six experiments comprise a factorial design, Shapley decompositions of mechanisms and moderators, a global sensitivity analysis over 600 parameter draws, a fixed-demand robustness variant, and a staggered-adoption panel of 5{,}000 simulated teams with known counterfactuals.
\textbf{Results.} \AGC{} raised change failure rate, rework and recovery time in both capability profiles. When AI raised change volume, it shortened lead time only where review capacity was spare ($-13\%$) and lengthened it elsewhere, sharply once review saturated; in the low-capability team its entire deployment-frequency gain was unplanned rework. \AGT{} improved deployment frequency ($+31$ to $+34\%$), lead time ($-8$ to $-22\%$) and recovery time ($-29$ to $-38\%$), but its stability effect depended on how often autonomous CI repair masked defects, and it raised escaped defects in the high-capability team. Capability reduced the absolute failure-rate penalty of \AGC{} (7.9 vs.\ 1.5 points) but not the relative one ($+34\%$ vs.\ $+84\%$). In the simulated panel, two-way fixed effects underestimated adoption effects by 8--16\%; detecting the modelled stability effect required about 50 teams, and its moderation by capability about 400.
\textbf{Conclusions.} Within the model, DORA metrics respond to AI through queueing, batching and oversight, not only through code quality. We derive measurement rules, guardrails for agentic remediation and a protocol, with sample-size requirements, for field validation.
\end{abstract}


\section{Introduction}\label{sec:intro}

Large language models are rapidly changing not only how software is written, but how software delivery systems operate. Within only a few years, AI-enabled development has progressed from passive inline code completion to increasingly autonomous agents capable of interpreting issues, modifying and testing code, opening pull requests, diagnosing failing builds, and responding to production telemetry \citep{jimenez2024swebench,yang2024sweagent,li2025rise}. This progression marks an important shift: AI is no longer confined to improving individual developer productivity, but is beginning to intervene directly in the socio-technical processes through which software changes are built, reviewed, deployed, recovered, and maintained. For organizations investing in these capabilities, the critical question is therefore no longer simply whether AI helps developers write code faster, but whether it improves software delivery performance at the system level. The de facto framework for evaluating such performance is the DORA family of metrics, including deployment frequency, change lead time, change failure rate, failed deployment recovery time, and, since 2024, deployment rework rate \citep{forsgren2018accelerate,dora2026metricsguide,dora2025metricshistory}.

The empirical evidence, however, remains fragmented and internally inconsistent. The 2024 DORA survey reported that a 25\% increase in AI adoption was associated with an estimated 1.5\% reduction in delivery throughput and a 7.2\% decline in delivery stability, despite respondents simultaneously reporting improvements in documentation quality, code quality, and review speed; the report proposed larger AI-enabled change batches as one possible explanation \citep{dora2024accelerate}. By 2025, the pattern had shifted: AI adoption was positively associated with throughput but remained negatively associated with stability, leading DORA to characterize AI as an \emph{amplifier} of an organization's existing engineering strengths and weaknesses rather than an independently beneficial capability \citep{dora2025aiassisted,dora2025aicapabilities}. Evidence from controlled productivity studies is similarly contradictory. In one experiment, developers completed a bounded programming task 55.8\% faster with an AI pair programmer \citep{peng2023impact}; in another, experienced open-source maintainers working within repositories they knew well were 19\% slower when using early-2025 AI tools, despite perceiving themselves to be faster \citep{becker2025measuring}. More importantly, most existing studies evaluate individual coding productivity rather than end-to-end software delivery performance. Direct empirical evidence on DORA outcomes remains exceptionally limited: one of the most complete published studies examines only 42 changes across two sprints within a single development team \citep{gunawan2026assessing}. Consequently, it remains unclear whether AI accelerates delivery, destabilizes it, or produces both effects through different mechanisms and under different organizational conditions.

Two conceptual problems make the evidence hard to accumulate. First, \emph{AI-generated code} and \emph{agentic AI} are usually merged into a single ``AI adoption'' variable although they act on different parts of the value stream. AI-generated code is a property of the change artifact and acts mainly on authoring, change size, defect content and review. Agentic AI is a property of the actor and can act anywhere in the lifecycle, including test generation, build repair, pipeline orchestration, incident diagnosis and rollback. The two can plausibly move the same metric in opposite directions. Second, DORA metrics are ratios and medians computed over deployments, so they react to changes in batching and deployment cadence as well as to changes in underlying quality. A metric can improve because its denominator grew; a speed gain at one stage can be absorbed by a queue at the next \citep{little1961proof,reinertsen2009principles}. Without an explicit model of the delivery system, observational associations cannot distinguish these cases.

This paper addresses both problems with a mechanism-based study. We formalize a conceptual model in which AI-generated code (\AGC{}) and agentic AI (\AGT{}) act on the value stream through distinct, literature-grounded mechanisms, moderated by four engineering capabilities. We implement the model as a stochastic discrete-event simulation of a team's delivery pipeline and use it as a computational laboratory \citep{davis2007developing,law2015simulation}: to derive the separate and joint effects implied by current evidence, to decompose them into mechanisms, to test when a speed--stability trade-off appears, to quantify how robust each conclusion is to parameter uncertainty, and to evaluate how well standard observational designs would recover the effects from field telemetry. The overarching question is:

\begin{quote}
\emph{How do agentic AI and AI-generated code individually and jointly affect DevOps operational performance, measured by deployment frequency, change lead time, change failure rate, failed deployment recovery time and deployment rework rate?}
\end{quote}

We decompose it into five research questions (\cref{sec:model}): the main effects of each construct (RQ1), their joint effect (RQ2), the existence and moderation of a speed--stability trade-off (RQ3), the mechanisms that transmit the effects (RQ4), and the dynamics after adoption together with their identifiability from field data (RQ5).

The paper makes four contributions.
\begin{enumerate}[label=(\roman*)]
\item \textbf{A graded evidence synthesis and a two-construct model.} We audit the quantitative claims in circulation, retain only those traceable to primary sources, and derive a conceptual model that separates \AGC{} from \AGT{} and links both to all five DORA metrics through explicit mechanisms and moderators (\cref{sec:background,sec:model}).
\item \textbf{A formal, open simulation model of AI-augmented delivery.} The model covers authoring, review queueing, continuous integration, deployment trains, production failure, latent defects, rework and recovery, and computes the five DORA metrics from first principles (\cref{sec:method}).
\item \textbf{Model-derived findings with quantified robustness.} We show that \AGC{}, when it raises change volume, shortens lead time only where review capacity can absorb the additional work; that it inflates deployment frequency with rework where capability is low; and that it raises instability under both capability profiles, whereas \AGT{} improves throughput and recovery but can \emph{increase} escaped defects in teams whose test suites are already strong. Capability dampens AI's harm in absolute terms but not in relative terms, which qualifies the ``amplifier'' proposition. Each directional hypothesis is accompanied by the share of 600 parameter draws in which it holds (\cref{sec:results}).
\item \textbf{A measurement and identification protocol with power analysis.} Using simulated teams with known counterfactuals, we show which estimators recover adoption effects from staggered rollouts, how many teams are needed to detect effects on stability and their moderation by capability, and which operational definitions the empirical follow-up requires (\cref{sec:results-design,sec:implications,app:protocol}).
\end{enumerate}

Throughout, we separate three kinds of statements: \emph{empirical evidence} from the literature, \emph{model-derived results} (properties of the simulated system under stated assumptions, estimated with Monte Carlo error), and \emph{interpretation and recommendations}. The simulation does not replace field measurement. It makes explicit which effects follow from the mechanisms the literature proposes, which depend on uncertain parameters, and what field data would be needed to tell them apart.

The remainder of the paper is organized as follows. \Cref{sec:background} reviews DORA metrics and the evidence on AI and delivery performance. \Cref{sec:model} presents the conceptual model, research questions and hypotheses. \Cref{sec:method} specifies the simulation model, its parameterization and the experiments. \Cref{sec:results} reports results by research question. \Cref{sec:discussion} discusses the findings and their implications, \cref{sec:threats} addresses threats to validity, and \cref{sec:conclusion} concludes.

\section{Background and Related Work}\label{sec:background}

\subsection{DORA software delivery metrics}\label{sec:dora}

DORA's research program established a small set of outcome metrics that jointly characterize software delivery performance and predict organizational performance \citep{forsgren2018accelerate}. The current definitions distinguish \emph{throughput}, comprising change lead time, deployment frequency and failed deployment recovery time, from \emph{instability}, comprising change failure rate and deployment rework rate \citep{dora2026metricsguide,dora2025metricshistory}. Two changes relative to the original four metrics matter here. First, \emph{failed deployment recovery time} (FDRT) replaced mean time to restore (MTTR): it measures ``the time it takes to recover from a deployment that fails and requires immediate intervention'' \citep{dora2026metricsguide}, and so excludes outages unrelated to changes. Second, \emph{deployment rework rate}, introduced in 2024, is ``the ratio of deployments that are unplanned but happen as a result of an incident in production'' \citep{dora2026metricsguide}. Rework rate captures defects that escape into production without triggering an immediate deployment failure, which is exactly the pathway through which latent quality problems in AI-generated code would surface. We use the current terminology and grouping throughout.

\subsection{AI-generated code and agentic AI as distinct constructs}\label{sec:constructs}

We define \emph{AI-generated code} (\AGC{}) as source code, tests or configuration whose content was substantially produced by a language model, whether by completion, chat or an agent, and whether or not a human edited it afterwards. \AGC{} is an attribute of a change. We define \emph{agentic AI} (\AGT{}) as AI systems that plan and execute multi-step actions with tool access (repositories, CI/CD systems, observability data, runbooks) with limited per-step human direction \citep{nisa2026agentic,hosseini2025role}. \AGT{} is an attribute of the actor that performs a lifecycle activity. Coding agents contribute to both constructs because they author changes; test-generation, build-repair, pipeline and incident-response agents contribute to \AGT{} without authoring production code. Keeping the constructs separate lets us test main effects and their interaction, and it matches how organizations adopt the tools: code assistants and coding agents are often introduced independently of agentic automation in operations \citep{sivakumar2024agentic,adabara2025trustworthy}.

\subsection{Evidence on AI and delivery performance}\label{sec:evidence}

\Cref{tab:evidence} summarizes the quantitative evidence we could trace to primary sources, grouped by design. Three observations motivate this study.

First, \emph{outcome-level evidence is thin}. The industry surveys are large but cross-sectional and self-reported, and they measure a single undifferentiated AI-adoption construct \citep{dora2024accelerate,dora2025aiassisted}. The only published quasi-experiment with all four classic metrics (F1000Research, which uses open post-publication peer review) reports a 39.2\% shorter mean lead time, a rise in deployment frequency from 2.1 to 3.4 per week, a fall in change failure rate from 14.3\% to 8.7\%, and recovery time falling from 4.2 to 2.8 hours. It rests on 42 changes over two sprints in one team, however \citep{gunawan2026assessing}. Our own earlier work on agentic AI in DevOps reports shorter test cycles and fewer rollbacks, but not the full metric set \citep{akour2025agentic}. Other contributions propose architectures or report improvements without effect sizes, or report them from short simulated environments \citep{brahmandam2025aiaugmented,sivakumar2024agentic,sugianto2025redefining}.

Second, \emph{repository-mining studies document mechanisms rather than outcomes}. Agentic pull requests are accepted less often than human ones: 35--64\% depending on the agent, against 76.8\% for human pull requests in the AIDev dataset \citep{li2025rise}. Among 567 Claude Code pull requests, 83.8\% were merged but 45.1\% of the merged ones required human revision \citep{watanabe2026agentic}. Each additional failed CI check lowers the odds that an agentic pull request is merged by about 15\% \citep{ehsani2026where}. AI-authored commits introduce issues, mostly code smells, of which 22.7\% survive to the latest repository version \citep{liu2026debt}. AI-generated pull requests show more redundancy and less reuse, yet reviewers respond to them more positively than to human ones \citep{huang2026more}. Industry code-quality data show copy-pasted lines rising and moved (refactored) lines falling in 2024 \citep{harding2025gitclear}. Earlier work found about 40\% of programs generated in security-relevant scenarios to be vulnerable \citep{pearce2022asleep}. These results suggest mechanisms: larger and more redundant changes, higher defect content, a review burden that reviewers may not perceive, and a technical-debt pathway into rework \citep{moreschini2026evolution,adalsteinsson2025rethinking}. None of them measures DORA outcomes.

Third, \emph{task-level productivity does not translate mechanically into delivery performance}. Speed-ups on bounded tasks \citep{peng2023impact,ziegler2024measuring} and slow-downs on familiar codebases \citep{becker2025measuring} both concern authoring time. DORA's change lead time, however, starts at commit. In a queueing system, faster authoring raises the arrival rate at review and deployment and lengthens queues as utilization approaches capacity \citep{little1961proof,kingman1961single,reinertsen2009principles}. Whether AI shortens or lengthens lead time is therefore a property of the whole delivery system, not of the tool.

\begin{table}[!tbp]
\centering
\caption{Quantitative evidence on AI and software delivery traced to primary sources, grouped by design. Grade reflects design strength for inferring effects on DORA outcomes (A: controlled or quasi-experimental with DORA outcomes; B: large-sample association with DORA constructs; C: mechanism evidence without DORA outcomes; D: simulated environment).}
\label{tab:evidence}
\small
\begin{tabularx}{\textwidth}{@{}p{3.1cm}p{3.2cm}Lc@{}}
\toprule
Source & Design / sample & Principal quantitative finding & Grade \\
\midrule
\citet{gunawan2026assessing} & Within-team quasi-experiment; 42 changes, 2 sprints & Lead time $-39.2\%$; DF 2.1$\to$3.4/week; CFR 14.3\%$\to$8.7\%; recovery 4.2$\to$2.8\,h & A \\
\citet{dora2024accelerate} & Cross-sectional practitioner survey & $+25\%$ AI adoption: throughput $-1.5\%$, stability $-7.2\%$ & B \\
\citet{dora2025aiassisted} & Survey, $\approx$5{,}000 respondents & AI adoption positively associated with throughput, negatively with stability; ``amplifier'' framing & B \\
\citet{li2025rise} & 456{,}535 agentic PRs & Acceptance 35--64\% (by agent) vs.\ 76.8\% for human PRs & C \\
\citet{watanabe2026agentic} & 567 agentic PRs, 157 projects & 83.8\% merged; 45.1\% of merged PRs revised by humans & C \\
\citet{ehsani2026where} & $\approx$33k agentic PRs & Each additional failed CI check: merge odds $\approx-15\%$ & C \\
\citet{liu2026debt} & 302.6k AI-authored commits, 6{,}299 repos & 22.7\% of AI-introduced issues survive to latest version & C \\
\citet{huang2026more} & AI vs.\ human PRs & More redundancy, less reuse; reviewer sentiment neutral/positive & C \\
\citet{harding2025gitclear} & 211M changed lines, 2020--2024 & Copy-pasted lines 10.6\%$\to$12.3\%, moved lines 15.9\%$\to$9.5\% (2023$\to$2024) & C \\
\citet{becker2025measuring} & RCT; 16 developers, 246 tasks & Completion time $+19\%$ with AI; developers estimated $-20\%$ & C \\
\citet{peng2023impact} & RCT; bounded task & Task completed 55.8\% faster & C \\
\citet{pearce2022asleep} & 1{,}689 generated programs & $\approx$40\% vulnerable & C \\
\citet{brahmandam2025aiaugmented} & 10-day simulated environment & DF, CFR and MTTR improvements under LLM agents & D \\
\bottomrule
\end{tabularx}
\end{table}

\subsection{Positioning}\label{sec:positioning}

Prior work leaves five gaps: (i) \AGC{} and \AGT{} are conflated; (ii) FDRT and rework rate are rarely measured; (iii) designs are cross-sectional or short before/after comparisons without controls; (iv) mechanisms such as batch size and review queuing are asserted rather than modelled; and (v) the amplifier proposition has not been expressed as a testable interaction. Our earlier work proposed measurement traceability from value-stream waste to goal-derived DORA metrics \citep{alenezi2026auditable} and model-driven derivation of pipelines from architecture decisions \citep{alenezi2026unified}. The present study supplies the causal-mechanism layer those frameworks lack for AI adoption: an explicit model of how AI changes flow through a pipeline and how that flow appears in DORA metrics. Simulation is established for building theory where empirical data are scarce and mechanisms interact nonlinearly \citep{davis2007developing}. We use it to sharpen hypotheses and to design the empirical study, not to replace one.

\section{Conceptual Model, Research Questions and Hypotheses}\label{sec:model}

\begin{figure}[!tbp]
\centering
\resizebox{\textwidth}{!}{%
\begin{tikzpicture}[
  font=\small,
  iv/.style={draw=black!70, fill=blue!8, rounded corners=3pt, minimum width=3.0cm, minimum height=1.0cm, align=center, thick},
  mech/.style={draw=black!50, fill=black!3, rounded corners=2pt, minimum width=3.6cm, minimum height=0.55cm, align=center},
  outn/.style={draw=black!70, rounded corners=2pt, minimum width=3.3cm, minimum height=0.6cm, align=center},
  grp/.style={draw=black!40, dashed, rounded corners=4pt, inner sep=5pt},
  arr/.style={-{Stealth[length=2.2mm]}, black!65, thick},
  marr/.style={-{Stealth[length=2.2mm]}, black!45, densely dashed}
]
\node[iv] (agc) at (0,1.3) {\textbf{AI-generated code}\\ share of changes $g$};
\node[iv] (agt) at (0,-1.6) {\textbf{Agentic AI}\\ lifecycle intensity $a$};

\node[mech] (m1) at (5.0,2.55) {Authoring speed-up (arrival rate)};
\node[mech] (m2) at (5.0,1.85) {Change-size inflation};
\node[mech] (m3) at (5.0,1.15) {Defect content};
\node[mech] (m4) at (5.0,0.45) {Review tax \& detection loss};
\node[mech] (m5) at (5.0,-0.45) {Agentic testing \& CI repair};
\node[mech] (m6) at (5.0,-1.15) {Pipeline acceleration};
\node[mech] (m7) at (5.0,-1.85) {Agentic diagnosis \& rollback};
\node[mech] (m8) at (5.0,-2.55) {Defect masking by agent fixes};
\begin{scope}[on background layer]
\node[grp, fit=(m1)(m8), label={[font=\small\bfseries]above:Mechanisms}] (G) {};
\end{scope}

\node[outn, fill=green!6] (df) at (10.3,1.9) {Deployment frequency $\uparrow$};
\node[outn, fill=green!6] (clt) at (10.3,1.15) {Change lead time $\downarrow$};
\node[outn, fill=green!6] (fdrt) at (10.3,0.4) {Recovery time (FDRT) $\downarrow$};
\node[outn, fill=red!6] (cfr) at (10.3,-0.85) {Change failure rate $\downarrow$};
\node[outn, fill=red!6] (drr) at (10.3,-1.6) {Deployment rework rate $\downarrow$};
\begin{scope}[on background layer]
\node[grp, fit=(df)(fdrt), label={[font=\small\bfseries]above:Throughput}] (T) {};
\node[grp, fit=(cfr)(drr), label={[font=\small\bfseries]below:Instability}] (I) {};
\end{scope}

\node[draw=black!60, fill=orange!8, rounded corners=3pt, align=center, minimum width=11.2cm, minimum height=0.7cm] (mod) at (5.3,-3.85)
 {\textbf{Moderators (engineering capabilities):} test automation $\cdot$ review capacity $\cdot$ small batches $\cdot$ platform \& observability};

\foreach \m in {m1,m2,m3,m4} \draw[arr] (agc.east) -- (\m.west);
\foreach \m in {m5,m6,m7,m8} \draw[arr] (agt.east) -- (\m.west);
\draw[arr] (G.east |- T.west) -- (T.west);
\draw[arr] (G.east |- I.west) -- (I.west);
\draw[marr] (agt.north) -- node[right, font=\footnotesize, black!60]{$g\times a$} (agc.south);
\draw[marr] (mod.north -| G.south) -- (G.south);
\draw[marr] (mod.north -| I.south) -- (I.south);
\end{tikzpicture}}
\caption{Conceptual model. AI-generated code (\AGC{}) and agentic AI (\AGT{}) affect the five DORA metrics through distinct mechanisms, jointly (interaction $g\times a$) and conditional on engineering capabilities. Arrows denote hypothesised causal paths; dashed arrows denote interaction and moderation. Throughput and instability follow DORA's current grouping \citep{dora2026metricsguide}.}
\label{fig:model}
\end{figure}
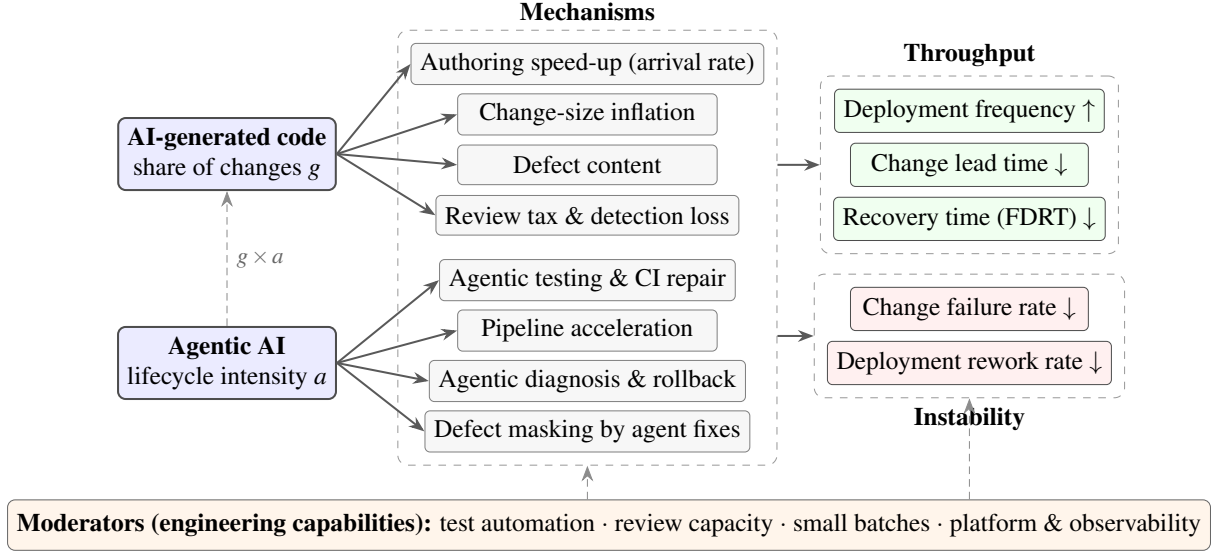

\Cref{fig:model} depicts the conceptual model. \AGC{} and \AGT{} act on the delivery system through mechanisms that change arrival rates, change size, defect content, review effort, detection, deployment cadence and recovery. These mechanisms determine the five DORA outcomes. Four engineering capabilities moderate the paths: \emph{test automation}, \emph{review capacity}, \emph{working in small batches}, and \emph{platform and observability} (deployment automation, detection and rollback). The last two correspond to capabilities in the DORA AI Capabilities Model \citep{dora2025aicapabilities}; the first two are the verification capacities most directly loaded by AI-generated changes.

The research questions follow from the gaps identified in \cref{sec:positioning}:
\begin{description}[leftmargin=1.2cm,style=nextline]
\item[RQ1] What are the separate effects of \AGC{} and \AGT{} on each DORA metric?
\item[RQ2] Do \AGC{} and \AGT{} interact, for example with agentic validation offsetting instability induced by AI-generated code?
\item[RQ3] Does AI adoption produce a speed--stability trade-off, and do engineering capabilities determine whether it appears (the amplifier proposition)?
\item[RQ4] Through which mechanisms are the effects transmitted?
\item[RQ5] How do effects evolve after adoption, and can standard observational designs recover them from team-level telemetry?
\end{description}

\Cref{tab:hypotheses} states the hypotheses, derived from the evidence in \cref{sec:evidence}. Each is a directional claim about a population of teams. In this paper we evaluate them \emph{within the model}. A hypothesis is \emph{model-supported} if the effect has the predicted sign at the calibrated parameter values and in a large majority of parameter draws. It is \emph{conditional} if the sign depends on capability or on uncertain parameters, and \emph{not supported} otherwise. Model support means the hypothesis follows from the mechanisms the literature proposes; it is not an empirical confirmation.

\begin{table}[!tbp]
\centering
\caption{Hypotheses. Signs refer to the metric's natural scale (e.g., ``$+$'' for CFR means more failures).}
\label{tab:hypotheses}
\small
\begin{tabularx}{\textwidth}{@{}lLcl@{}}
\toprule
ID & Hypothesis & Sign & RQ \\
\midrule
H1a & \AGC{} increases deployment frequency. & $+$ & RQ1 \\
H1b & \AGT{} increases deployment frequency. & $+$ & RQ1 \\
H2a & \AGC{} shortens coding but lengthens review waiting; its net effect on change lead time depends on review capacity. & $\pm$ & RQ1, RQ3 \\
H2b & \AGT{} shortens change lead time. & $-$ & RQ1 \\
H3a & \AGC{} increases change failure rate. & $+$ & RQ1 \\
H3b & \AGT{} applied to testing and validation decreases change failure rate. & $-$ & RQ1 \\
H3c & \AGT{} attenuates the CFR penalty of \AGC{} (negative interaction). & $-$ & RQ2 \\
H4a & \AGT{} applied to operations shortens failed deployment recovery time. & $-$ & RQ1 \\
H4b & \AGC{} lengthens failed deployment recovery time (harder diagnosis, larger deployments). & $+$ & RQ1 \\
H5 & \AGC{} increases deployment rework rate. & $+$ & RQ1 \\
H6 & Engineering capabilities moderate H1--H5 such that high-capability teams gain throughput without stability loss (amplifier). & mod. & RQ3 \\
H7 & Change size and review latency transmit (mediate) the effects of \AGC{} on CLT and CFR. & med. & RQ4 \\
H8 & Stability effects follow a J-curve: initial deterioration after adoption, then partial recovery. & dyn. & RQ5 \\
\bottomrule
\end{tabularx}
\end{table}

\section{Method}\label{sec:method}

\subsection{Research design}\label{sec:design}

We follow a model-based design in three steps \citep{davis2007developing,sargent2013verification}. (1) We translate the conceptual model into a stochastic discrete-event simulation of one team's delivery pipeline, with each mechanism in \cref{fig:model} implemented as an explicit, switchable rule. (2) We run six computational experiments (E1--E6, \cref{sec:experiments}) that map onto the research questions. (3) We analyze the outputs with statistical procedures that mirror those an empirical study would use: regression with robust standard errors, Shapley decomposition, global sensitivity analysis, and staggered difference-in-differences with bootstrap inference. Everything is implemented in Python (NumPy, pandas, statsmodels).

\subsection{Simulation model}\label{sec:sim}

\paragraph{Time and team.} Time is measured in working hours, with 40 working hours per week. A team of $n_d=8$ developers produces changes, $n_r$ reviewers review them in FIFO order, and a deployment train releases merged changes to production. Each run simulates a 4-week warm-up followed by $n_w=52$ observed weeks. Developers work from an unbounded backlog: each starts a new change as soon as the previous one is authored. Faster authoring therefore raises the number of changes; \cref{sec:robustness} relaxes this assumption.

\paragraph{Adoption.} AI adoption is described by two scalars in $[0,1]$: the \AGC{} share $g$, the probability that a change is AI-generated, and the \AGT{} intensity $a$. For CI repair, $a$ is the probability that an agent performs the fix; for the other agentic mechanisms it scales the mechanism's effect linearly, reaching the full effect $\eta_\bullet$ at $a=1$. In staggered-adoption experiments both ramp up after the adoption time $t_0$ as $r(t)=1-e^{-(t-t_0)/T_r}$ with $T_r=4$ weeks.

\paragraph{Capabilities.} Four capability levels $\kappa_{\text{test}},\kappa_{\text{rev}},\kappa_{\text{batch}},\kappa_{\text{plat}}\in[0,1]$ interpolate each capability-dependent parameter between a low and a high endpoint (\cref{tab:params}): linearly, except the deployment interval, which is interpolated on the log scale. \emph{Low-capability} and \emph{high-capability} teams set all four to 0 and 1 respectively.

\paragraph{Authoring.} Developers author changes sequentially. Change $i$ is AI-generated with indicator $A_i\sim\Bern(g\,r(t))$. Its size (lines of code), authoring time and number of injected defects are
\begin{align}
L_i &= s\,L_0\,e^{\sigma_L Z_i}\,\bigl[1+(\lambda_L-1)(1-\delta)\bigr]^{A_i}, \qquad Z_i\sim\mathcal N(0,1), \label{eq:size}\\
T^{\text{auth}}_i &\sim \LN\!\bigl(\text{mean}=m_A\,s^{0.7}(1-\lambda_S)^{A_i},\ \text{cv}=0.8\bigr), \label{eq:auth}\\
D_i &\sim \Pois\!\bigl(\theta_D\,\lambda_D^{A_i}\,L_i/100\bigr), \label{eq:defects}
\end{align}
where $s$ is the change-size scale (small-batch discipline), $\delta$ damps AI size inflation, and $\lambda_S$, $\lambda_L$ and $\lambda_D$ are the \AGC{} speed-up, size and defect multipliers. The first commit occurs at a fraction $\varphi=0.5$ of authoring time. This matters because DORA's lead time starts at commit, so only part of any authoring speed-up can shorten it.

\paragraph{Review.} Completed changes join a FIFO queue served by $n_r$ reviewers. Review time is
\begin{equation}
S_i=s_R\,(L_i/100)^{0.7}\,\lambda_R^{A_i}\,\epsilon_i, \label{eq:service}
\end{equation}
where $\epsilon_i$ is lognormal with mean 1 and log-scale standard deviation 0.5, and $\lambda_R$ is the verification tax on AI-generated changes. Each defect is detected in review with probability
\begin{equation}
p^{\text{rev}}_i = p_0\,\min\!\bigl(1,(100/L_i)^{1/2}\bigr)\,\lambda_P^{A_i}, \label{eq:review}
\end{equation}
which declines with change size and, through $\lambda_P<1$, for AI-generated changes that ``look clean'' \citep{huang2026more}. Detected defects are fixed by the author and re-checked outside the reviewer queue.

\paragraph{Continuous integration.} The test suite detects each remaining defect with probability
\begin{equation}
\tau_{\text{eff}} = 1-(1-\tau)(1-\eta_T\,a), \label{eq:tau}
\end{equation}
where $\tau$ is test-suite strength and $\eta_T$ is the share of test-escaping defects that agent-generated tests catch at full intensity. Undetected defects escape because the suite is deterministic. A CI failure, whether from detected defects or a flaky run with probability $0.05$, triggers a fix. With probability $a$ an agent performs the fix, which is faster by a factor $(1-\eta_F)$ but \emph{masks} each detected defect with probability $\mu$ (for example by weakening a test) instead of repairing it, so the defect escapes. This masking term captures the oversight risk of autonomous remediation \citep{adabara2025trustworthy}.

\paragraph{Deployment.} Merged changes wait for the next deployment-train departure. Trains depart every $\Delta(1-\eta_P a)$ hours, and each carries all waiting changes, so deployment batch size emerges from cadence and arrival rate rather than being imposed.

\paragraph{Production.} Each escaped defect in a deployment manifests immediately with probability $q$. If at least one does, the deployment \emph{fails}. Otherwise the defect is latent: with probability $p_s$ it surfaces as a user-facing bug after an $\Expo$ time (mean 80\,h) and is corrected by an unplanned hotfix deployment, which counts as rework.

\paragraph{Recovery.} A failed deployment $k$ carrying $n_k$ changes is recovered after $R_k=T^{\text{det}}_k+T^{\text{diag}}_k+T^{\text{rem}}_k$, where
\begin{equation}
\E[T^{\text{diag}}_k] = d_0\,\bigl(1+\beta_B(n_k-1)\bigr)\,(1+\lambda_G)^{\1[\text{AI cause}]}\,(1-\eta_O a). \label{eq:diag}
\end{equation}
Diagnosis thus slows with the number of changes in the deployment \citep{reinertsen2009principles}, with an AI-generated cause ($\lambda_G$), and speeds up with agentic diagnosis support ($\eta_O$). Remediation is a rollback (0.3\,h) with probability $\pi+(1-\pi)\eta_{RB}\,a$. Otherwise it is a fix-forward, which requires an unplanned rework deployment.

\paragraph{Outcome metrics.} Let $\mathcal K$ be the set of deployments in the observation window, $\mathcal K_F\subseteq\mathcal K$ the failed ones, $\mathcal K_R\subseteq\mathcal K$ the unplanned rework deployments, and $\mathcal C$ the deployed changes. Then
\begin{gather}
\DF=\frac{|\mathcal K|}{n_w},\qquad
\CFR=\frac{|\mathcal K_F|}{|\mathcal K|},\qquad
\DRR=\frac{|\mathcal K_R|}{|\mathcal K|}, \nonumber\\
\CLT=\median_{i\in\mathcal C}\bigl(t^{\text{dep}}_i-t^{\text{commit}}_i\bigr),\qquad
\FDRT=\median_{k\in\mathcal K_F}R_k. \label{eq:metrics}
\end{gather}
These follow the DORA definitions \citep{dora2026metricsguide}. We add two diagnostic metrics: \emph{planned deployment frequency} $|\mathcal K\setminus\mathcal K_R|/n_w$, which excludes rework, and \emph{failure intensity} $\FI=|\mathcal K_F|/n_w$, the number of failed deployments per week. $\CFR$ can fall simply because $|\mathcal K|$ grows, and $\DF$ can rise simply because rework grows. The model also records mechanism variables: change size, review waiting time, first-run CI failure rate, escaped defects per change and changes per deployment. \Cref{alg:sim} summarizes one simulation run.

\begin{algorithm}[!tbp]
\caption{One simulation run (one team, one scenario).}\label{alg:sim}
\small
\begin{algorithmic}[1]
\Require adoption $(g,a)$, capabilities $\bm\kappa$, parameters $\bm\theta$, seed
\For{each developer} \Comment{authoring, \cref{eq:size,eq:auth,eq:defects}}
  \State generate successive changes with $A_i$, $L_i$, $T^{\text{auth}}_i$, $D_i$ until the horizon
\EndFor
\State sort changes by completion time; serve with $n_r$ FIFO reviewers; draw review detections (\cref{eq:review})
\For{each change} \Comment{CI, \cref{eq:tau}}
  \State detect defects with $\tau_{\text{eff}}$; on failure, fix (agent w.p.\ $a$: faster, masks w.p.\ $\mu$); record escapes
\EndFor
\State release all merged changes at each train departure (interval $\Delta(1-\eta_P a)$)
\For{each deployment}
  \State split escaped defects into immediate and latent; schedule hotfix rework for surfaced latent defects
  \If{any immediate defect} record failure; draw $R_k$ (\cref{eq:diag}); rollback or fix-forward (rework) \EndIf
\EndFor
\State \Return metrics (\cref{eq:metrics}) and mechanism variables over weeks $5$--$56$
\end{algorithmic}
\end{algorithm}

\subsection{Parameterisation and calibration}\label{sec:params}

\Cref{tab:params} lists all parameters. The AI mechanism parameters are informed by the evidence of \cref{tab:evidence} where related evidence exists; none of them has been measured directly, so all are varied over wide ranges in the global sensitivity analysis (E4). The ranges are deliberately broad. The \AGC{} speed-up range $[-0.2,0.6]$, for example, spans both the 19\% slow-down observed by \citet{becker2025measuring} and the 55.8\% speed-up of \citet{peng2023impact}. The \AGC{} defect multiplier range $[0.8,1.8]$ allows AI-generated changes to be \emph{less} defect-prone than human ones. Process parameters were calibrated to two contrasting but plausible teams (\cref{tab:levels}, baseline rows). The low-capability team releases on a schedule, deploys about 3.6 times per week (2.5 planned), has lead times of about two working days, recovers from failures within about a working day, and has a CFR above 20\%. Its throughput is therefore moderate and its stability poor. The high-capability team deploys several times per day, with lead times under one working day, recovery in about two hours and a CFR of about 2\%, comparable to the best-performing DORA profiles. These are plausibility anchors, not estimates for any real organization.

\begin{table}[!tbp]
\centering
\caption{Model parameters. The last column gives, for the AI mechanism parameters, the range sampled in the global sensitivity analysis (E4) and the evidence that informed the baseline value; none of these parameters has been measured directly.}
\label{tab:params}
\footnotesize
\begin{tabularx}{\textwidth}{@{}>{\raggedright\arraybackslash}p{1.9cm}>{\raggedright\arraybackslash}p{4.6cm}>{\centering\arraybackslash}p{2.3cm}L@{}}
\toprule
Symbol & Meaning & Baseline & E4 range / grounding \\
\midrule
\multicolumn{4}{@{}l}{\textit{AI-generated code (\AGC{})}}\\
$\lambda_S$ & authoring-time reduction & 0.40 & $[-0.2,0.6]$; \citet{becker2025measuring,peng2023impact} \\
$\lambda_L$ & change-size multiplier & 1.50 & $[1.0,2.0]$; assumption informed by \citet{huang2026more,dora2024accelerate} \\
$\lambda_D$ & defect-rate multiplier & 1.35 & $[0.8,1.8]$; assumption informed by \citet{li2025rise,watanabe2026agentic,pearce2022asleep} \\
$\lambda_R$ & review-time multiplier (verification tax) & 1.30 & $[1.0,1.6]$; assumption informed by \citet{adalsteinsson2025rethinking,watanabe2026agentic} \\
$\lambda_P$ & review detectability multiplier & 0.80 & $[0.6,1.0]$; assumption informed by \citet{huang2026more} \\
$\lambda_G$ & extra diagnosis time if AI-caused & 0.40 & $[0.0,0.8]$; assumption \\
\addlinespace
\multicolumn{4}{@{}l}{\textit{Agentic AI (\AGT{}), effects at $a=1$}}\\
$\eta_T$ & share of test escapes caught by agent tests & 0.40 & $[0.1,0.7]$; assumption \\
$\mu$ & masking probability of agent CI fixes & 0.15 & $[0.0,0.3]$; assumption; oversight risk \citep{adabara2025trustworthy} \\
$\eta_F$ & CI-fix time reduction & 0.75 & $[0.5,0.9]$; assumption informed by \citet{li2025rise} \\
$\eta_P$ & deployment-interval reduction & 0.50 & $[0.2,0.7]$; assumption \\
$\eta_O$ & diagnosis-time reduction & 0.40 & $[0.1,0.6]$; assumption informed by \citet{gunawan2026assessing} \\
$\eta_{RB}$ & share of fix-forwards made rollbackable & 0.50 & $[0.2,0.8]$; assumption informed by \citet{akour2025agentic} \\
\addlinespace
\multicolumn{4}{@{}l}{\textit{Process (fixed)}}\\
$n_d$, $m_A$ & developers; authoring mean (h), cv 0.8 & 8; 12 & -- \\
$L_0$, $\sigma_L$ & median change size (LOC); log-size s.d. & 80; 1.0 & -- \\
$\theta_D$ & defects per 100 LOC (human) & 0.30 & -- \\
$s_R$, $p_0$ & review h per 100 LOC; review detection (size exponents 0.7, 0.5) & 1.0; 0.35 & -- \\
-- & review-finding fix time (h) & Exp(2.0)$+0.3$ & -- \\
-- & CI run time (h); flaky-failure prob.\ (fix 0.1\,h); human CI-fix mean (h) & $0.5(1-0.5\kappa_{\text{plat}})$; 0.05; 1.2 & -- \\
$q$, $p_s$ & immediate-failure share; latent surfacing share (mean delay 80\,h) & 0.30; 0.25 & -- \\
$\beta_B$ & diagnosis slowdown per extra change in a deployment & 0.10 & -- \\
-- & diagnosis cv; rollback time (h); fix-forward (h); hotfix mean (h) & 0.8; 0.3; Exp(2.0)$+0.25$; 3.0 & -- \\
\addlinespace
\multicolumn{4}{@{}l}{\textit{Capabilities: (low, high) endpoints}}\\
$\tau$ & test-suite strength (test automation) & (0.55, 0.85) & -- \\
$n_r$ & reviewers (review capacity) & (2, 4) & -- \\
$s$, $\delta$ & size scale; AI size damping (small batches) & (1.0, 0.6); (0, 0.7) & -- \\
$\Delta$ & deployment interval, h (platform; log-interpolated) & (16, 2) & -- \\
$T^{\text{det}}$, $d_0$, $\pi$ & detection mean (h); diagnosis mean (h); rollback prob.\ (platform) & (1.0, 0.25); (4.0, 1.5); (0.4, 0.8) & -- \\
\bottomrule
\end{tabularx}
\end{table}

\subsection{Experiments and analysis}\label{sec:experiments}

\paragraph{E1: factorial experiment (RQ1--RQ3).} A full factorial over $g,a\in\{0,0.25,0.5,0.75,1\}$ and capability $\in\{\text{low},\text{high}\}$, with 100 replications per cell (5{,}000 runs). Within each capability profile we estimate
\begin{equation}
\ln Y = \beta_0+\beta_g\,g+\beta_a\,a+\beta_{ga}\,g a+\xi \label{eq:reg}
\end{equation}
by OLS with HC3 standard errors on cells with $g,a\le 0.75$; cells with $g=1$ are analyzed separately because they include a regime change (\cref{sec:rq1}). The coefficients are semi-elasticities: $\beta_g$ is the log change of the metric per unit of \AGC{} share when $a=0$, and $\beta_{ga}$ measures how that slope changes with $a$. The linear form is a summary; where responses are strongly nonlinear (lead time in the low-capability team) we also report cell contrasts. To study the trade-off (RQ3) we summarize each scenario by a throughput index and an instability index relative to the same team's no-AI baseline,
\begin{equation}
\Delta\mathrm{TI}=\tfrac13\Bigl[\ln\tfrac{\DF^{\text{pl}}}{\DF^{\text{pl}}_0}-\ln\tfrac{\CLT}{\CLT_0}-\ln\tfrac{\FDRT}{\FDRT_0}\Bigr],\qquad
\Delta\mathrm{II}=\tfrac12\Bigl[\ln\tfrac{\CFR}{\CFR_0}+\ln\tfrac{\DRR}{\DRR_0}\Bigr], \label{eq:indices}
\end{equation}
where $\DF^{\text{pl}}$ is planned deployment frequency, so that rework does not count as throughput. A scenario is ``faster'' if $\Delta\mathrm{TI}>0$ and ``less stable'' if $\Delta\mathrm{II}>0$.

\paragraph{E2: mechanism decomposition (RQ4).} We attribute the \AGC{} effect ($g=0.75$, $a=0$) to its six mechanisms $\mathcal M=\{$speed-up, size, defects, review tax, review detection, diagnosis$\}$ by running all $2^6=64$ on/off combinations with common random numbers (100 seeds each) and computing Shapley values \citep{shapley1953value}:
\begin{equation}
\phi_m=\sum_{S\subseteq\mathcal M\setminus\{m\}}\frac{|S|!\,(|\mathcal M|-|S|-1)!}{|\mathcal M|!}\bigl[v(S\cup\{m\})-v(S)\bigr], \label{eq:shapley}
\end{equation}
where $v(S)$ is the mean metric with only the mechanisms in $S$ active. The $\phi_m$ sum exactly to the total effect and, unlike one-at-a-time ablation, split interaction effects fairly. Confidence intervals come from 300 bootstrap resamples of seeds.

\paragraph{E3: moderator decomposition (RQ3).} We define the AI effect of a capability profile as the paired difference between $(g,a)=(0.75,0.5)$ and $(0,0)$, and compute its value for all $2^4$ combinations of low and high moderators (100 seeds each). Shapley values over the four moderators then attribute the change in the AI effect, from the all-low to the all-high team, to individual capabilities.

\paragraph{E4: global sensitivity analysis (robustness).} We draw 600 Latin-hypercube samples \citep{mckay1979comparison} of the twelve AI mechanism parameters over the ranges in \cref{tab:params}. For each draw and capability profile we simulate four arms, (baseline, \AGC{} only, \AGT{} only, both) at $g,a\in\{0,0.75\}$, with six common-random-number replications. From these we compute the relative \AGC{} and \AGT{} effects and their log-scale interaction. We report the share of draws in which each hypothesised sign holds, and standardised regression coefficients (SRC) of each effect on the parameters \citep{saltelli2008global}. Because each draw averages only six replications, these shares include some Monte Carlo noise in addition to parameter uncertainty.

\paragraph{E5: staggered adoption panel (RQ5 and design).} We simulate 5{,}000 heterogeneous teams. Each has a capability level $c\sim\mathrm{Beta}(2,2)$ with component-specific noise, an adoption cohort in weeks $\{16,24,32\}$ or never (30\%), and target adoption $g\sim U(0.4,0.9)$, $a\sim U(0,0.6)$. AI penalties decay after adoption through organisational learning with a time constant of $8+32(1-c)$ weeks to a residual of 50\%, and all teams share a secular trend and seasonality in defect rates. Each adopting team is also simulated without adoption using the same seed. Because every developer and pipeline stage draws from its own random stream, the two runs are identical up to adoption, so their difference is the team's \emph{true} effect. Outcomes are aggregated into thirteen 4-week blocks. We compare three estimators against the ground-truth average treatment effect on the treated (ATT): a naive pre/post difference for adopters; two-way fixed effects (TWFE), $Y_{jt}=\alpha_j+\gamma_t+\beta D_{jt}+\xi_{jt}$; and the \citet{callaway2021difference} estimator with never-adopting controls,
\begin{equation}
\mathrm{ATT}(G,t)=\E[Y_{t}-Y_{G-1}\mid G_j=G]-\E[Y_{t}-Y_{G-1}\mid G_j=\infty], \label{eq:cs}
\end{equation}
for post-adoption periods $t\ge G$ (pre-adoption placebo estimates use the preceding block $t-1$ as base), aggregated across cohorts and periods with cohort-size weights, or by event time $e=t-G$ for the dynamic analysis. Log deployment frequency is computed as $\ln\bigl((|\mathcal K_b|+0.5)/4\bigr)$ for block $b$. Confidence intervals use a team-level bootstrap (300 replications; 100 for TWFE). TWFE is included because it is biased under staggered adoption with dynamic effects \citep{goodmanbacon2021difference,sun2021estimating}. Power is estimated from 300 panels per size $N\in\{12,24,48,96,192,384\}$ teams sampled without replacement from this population, with 199-replicate team-level bootstrap confidence intervals.

\paragraph{E6: fixed-demand variant (robustness).} Because the unbounded-backlog assumption lets faster authoring raise change volume, we repeat the E1 anchor cells ($g,a\in\{0,0.75\}$, both capability profiles, 100 replications) with fixed demand: AI shortens the authoring of a change, but each developer starts changes at the human pace (\cref{app:fixeddemand}).

\paragraph{Verification.} We verified the implementation in four ways (\cref{app:verification}): invariance of the DORA outputs to \AGC{} parameters when $g=0$ and to \AGT{} parameters when $a=0$; a null-mechanism test (with all \AGC{} mechanisms disabled, the \AGC{} effect is exactly zero); identity of pre-adoption histories in the panel experiment; and qualitative agreement of simulated review waiting times with the Allen--Cunneen $M/G/c$ approximation \citep{allen1990probability}.

\section{Results}\label{sec:results}

All results in this section are model-derived. Reported intervals are 95\% confidence intervals for Monte Carlo estimates and reflect simulation noise, not uncertainty about the real world. That second kind of uncertainty is addressed by the sensitivity analysis in \cref{sec:robustness}.

\subsection{RQ1: separate effects of AI-generated code and agentic AI}\label{sec:rq1}

\begin{figure}[!tbp]
\centering
\includegraphics[width=\textwidth]{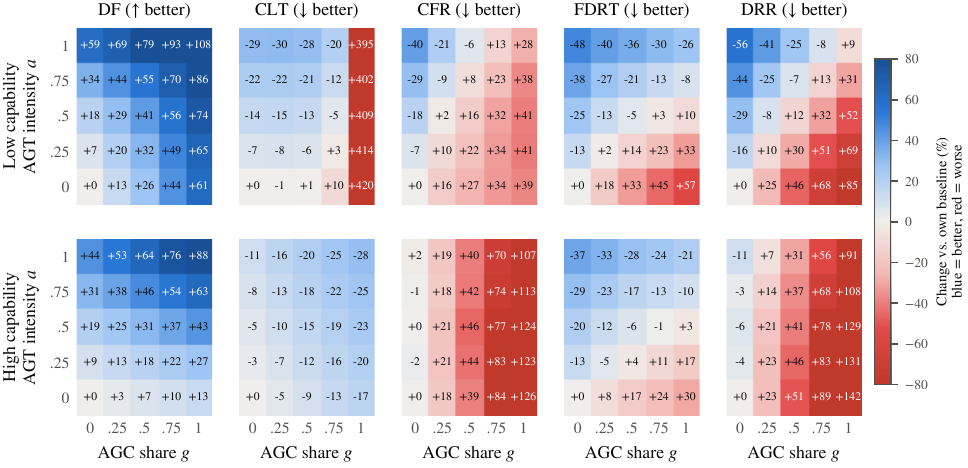}
\caption{Percentage change in each DORA metric relative to the same team's no-AI baseline across the E1 factorial (columns: \AGC{} share $g$; rows: \AGT{} intensity $a$). Color encodes desirability (blue = better, red = worse, saturating at $\pm80\%$); cell labels give the signed change. Each cell is the mean of 100 runs. DF follows the DORA definition and includes rework deployments; planned DF is reported in \cref{tab:levels}.}
\label{fig:heatmap}
\end{figure}

\Cref{tab:levels} reports steady-state DORA levels for four anchor scenarios, \cref{tab:regression} the semi-elasticities from \cref{eq:reg}, and \cref{fig:heatmap} the full factorial as percentage changes from each team's own no-AI baseline.

\begin{table}[!tbp]
\centering
\caption{Steady-state DORA metrics for anchor scenarios (E1; \AGC{} share $g$ and \AGT{} intensity $a$ set to 0.75 where active; means of 100 runs with 95\% bootstrap CIs). Planned DF excludes rework deployments; FI is failed deployments per week.}
\label{tab:levels}
\resizebox{\textwidth}{!}{%
\begin{tabular}{@{}lccccccc@{}}
\toprule
Scenario & DF (/week) & Planned DF (/week) & CLT (h) & CFR (\%) & FDRT (h) & DRR (\%) & FI (/week) \\
\midrule
\multicolumn{8}{@{}l}{\textit{Low capability team}}\\
\quad Baseline & 3.6 \scriptsize[3.6, 3.7] & 2.5 \scriptsize[2.5, 2.5] & 15.7 \scriptsize[15.7, 15.8] & 23.5 \scriptsize[22.9, 24.0] & 8.9 \scriptsize[8.7, 9.1] & 31.0 \scriptsize[30.5, 31.5] & 0.85 \scriptsize[0.83, 0.87] \\
\quad AGC only & 5.2 \scriptsize[5.2, 5.3] & 2.5 \scriptsize[2.5, 2.5] & 17.4 \scriptsize[17.3, 17.5] & 31.4 \scriptsize[31.0, 31.8] & 12.9 \scriptsize[12.7, 13.2] & 52.2 \scriptsize[51.7, 52.6] & 1.64 \scriptsize[1.62, 1.66] \\
\quad AGT only & 4.8 \scriptsize[4.8, 4.9] & 4.0 \scriptsize[4.0, 4.0] & 12.3 \scriptsize[12.3, 12.4] & 16.7 \scriptsize[16.3, 17.2] & 5.6 \scriptsize[5.4, 5.7] & 17.4 \scriptsize[16.9, 17.8] & 0.81 \scriptsize[0.79, 0.83] \\
\quad AGC + AGT & 6.2 \scriptsize[6.1, 6.2] & 4.0 \scriptsize[4.0, 4.0] & 13.9 \scriptsize[13.8, 14.0] & 28.9 \scriptsize[28.4, 29.3] & 7.7 \scriptsize[7.6, 7.9] & 35.1 \scriptsize[34.7, 35.5] & 1.78 \scriptsize[1.75, 1.81] \\
\addlinespace
\multicolumn{8}{@{}l}{\textit{High capability team}}\\
\quad Baseline & 17.8 \scriptsize[17.7, 17.8] & 17.5 \scriptsize[17.5, 17.6] & 5.6 \scriptsize[5.6, 5.7] & 1.8 \scriptsize[1.7, 1.8] & 2.3 \scriptsize[2.2, 2.4] & 1.3 \scriptsize[1.3, 1.4] & 0.31 \scriptsize[0.30, 0.33] \\
\quad AGC only & 19.5 \scriptsize[19.5, 19.5] & 19.0 \scriptsize[19.0, 19.0] & 4.9 \scriptsize[4.9, 4.9] & 3.2 \scriptsize[3.1, 3.4] & 2.8 \scriptsize[2.7, 2.9] & 2.5 \scriptsize[2.4, 2.6] & 0.63 \scriptsize[0.61, 0.65] \\
\quad AGT only & 23.3 \scriptsize[23.3, 23.4] & 23.0 \scriptsize[23.0, 23.1] & 5.2 \scriptsize[5.2, 5.2] & 1.7 \scriptsize[1.7, 1.8] & 1.6 \scriptsize[1.6, 1.7] & 1.3 \scriptsize[1.2, 1.4] & 0.41 \scriptsize[0.39, 0.42] \\
\quad AGC + AGT & 27.4 \scriptsize[27.4, 27.5] & 26.8 \scriptsize[26.8, 26.9] & 4.4 \scriptsize[4.4, 4.4] & 3.1 \scriptsize[3.0, 3.1] & 2.0 \scriptsize[1.9, 2.0] & 2.2 \scriptsize[2.2, 2.3] & 0.84 \scriptsize[0.82, 0.86] \\
 
\bottomrule
\end{tabular}}
\end{table}

\paragraph{AI-generated code.} In both capability profiles, \AGC{} raises deployment frequency, change failure rate, recovery time and rework (\cref{tab:levels}). How much of that is real throughput depends on the team. In the low-capability team, deployment frequency rises from 3.6 to 5.2 per week at $g=0.75$ ($+44\%$), yet \emph{planned} deployment frequency stays at 2.5 per week, fixed by the release cadence. The entire increase consists of unplanned rework deployments: DRR rises from 31.0\% to 52.2\%. Change lead time increases by 10\% (15.7 to 17.4\,h), because the authoring speed-up is offset by longer review queues. At $g=1$ review utilization reaches about one and the queue saturates. Mean review waiting grows from 2.8\,h to 72\,h and median lead time to 82\,h (\cref{fig:heatmap}, CLT column). Because 6.6\% of the changes authored in the window are still undeployed at the horizon, this median is a lower bound that grows with the length of the horizon. In the high-capability team, \AGC{} shortens lead time by 13\% (5.6 to 4.9\,h) and raises planned deployment frequency from 17.5 to 19.0 per week, but it also nearly doubles change failure rate (1.8\% to 3.2\%) and doubles failure intensity (0.31 to 0.63 failed deployments per week). Recovery time lengthens in both teams, by 45\% and 24\% respectively, because failed deployments carry more changes and AI-caused failures take longer to diagnose (\cref{sec:rq4}).

\paragraph{Agentic AI.} \AGT{} improves every throughput metric in both profiles: DF $+34\%$ and $+31\%$, CLT $-22\%$ and $-8\%$, FDRT $-38\%$ and $-29\%$ for the low- and high-capability teams. Its stability effect depends on capability. In the low-capability team, CFR falls from 23.5\% to 16.7\% and DRR from 31.0\% to 17.4\%. Failure intensity, however, falls by only 5\% (0.85 to 0.81 per week): most of the CFR improvement comes from more, smaller deployments (10.7 to 6.7 changes per deployment), not from fewer failures. In the high-capability team, CFR is unchanged (1.8\% vs.\ 1.7\%, \cref{tab:regression}: $\beta_a=+0.02$, n.s.), whereas escaped defects per change rise from 0.026 to 0.036 and failure intensity from 0.31 to 0.41 per week. A strong test suite catches most defects, so agentic CI repair has many detected defects to act on, and masking a share of them outweighs the extra defects that agent-generated tests catch; \cref{sec:robustness} shows that this sign depends on the masking rate. Because $a$ scales all agentic activities jointly in E1, the separate contributions of testing, pipeline and operations agents are identified only in the sensitivity analysis.

\begin{table}[!tbp]
\centering
\caption{Semi-elasticities from \cref{eq:reg} ($\ln Y$ on $g$, $a$ and $g\times a$, cells with $g,a\le0.75$, 1{,}600 runs per profile, HC3 95\% CIs). For $g$ and $a$, a coefficient $\beta$ implies a $100(e^{0.75\beta}-1)\%$ change when the regressor moves from 0 to 0.75 with the other regressor at 0 (linear approximation). $^\dagger$: not significant at $\alpha=0.05$.}
\label{tab:regression}
\resizebox{\textwidth}{!}{%
\begin{tabular}{@{}lcccccccccc@{}}
\toprule
& \multicolumn{2}{c}{DF} & \multicolumn{2}{c}{CLT} & \multicolumn{2}{c}{CFR} & \multicolumn{2}{c}{FDRT} & \multicolumn{2}{c}{DRR} \\
\cmidrule(lr){2-3}\cmidrule(lr){4-5}\cmidrule(lr){6-7}\cmidrule(lr){8-9}\cmidrule(lr){10-11}
Term / profile & $\beta$ & 95\% CI & $\beta$ & 95\% CI & $\beta$ & 95\% CI & $\beta$ & 95\% CI & $\beta$ & 95\% CI \\
\midrule
\multicolumn{11}{l}{\textit{AGC share $g$}}\\
\quad Low capability & +0.49 & \scriptsize[+0.48, +0.50] & +0.12 & \scriptsize[+0.11, +0.13] & +0.38 & \scriptsize[+0.35, +0.41] & +0.49 & \scriptsize[+0.46, +0.52] & +0.69 & \scriptsize[+0.67, +0.71] \\
\quad High capability & +0.12 & \scriptsize[+0.12, +0.12] & -0.19 & \scriptsize[-0.19, -0.19] & +0.83 & \scriptsize[+0.77, +0.89] & +0.32 & \scriptsize[+0.27, +0.36] & +0.91 & \scriptsize[+0.83, +0.99] \\
\addlinespace
\multicolumn{11}{l}{\textit{AGT intensity $a$}}\\
\quad Low capability & +0.39 & \scriptsize[+0.37, +0.40] & -0.33 & \scriptsize[-0.34, -0.32] & -0.45 & \scriptsize[-0.49, -0.42] & -0.63 & \scriptsize[-0.66, -0.59] & -0.76 & \scriptsize[-0.79, -0.73] \\
\quad High capability & +0.36 & \scriptsize[+0.36, +0.36] & -0.11 & \scriptsize[-0.12, -0.11] & +0.02$^{\dagger}$ & \scriptsize[-0.04, +0.08] & -0.43 & \scriptsize[-0.47, -0.38] & -0.02$^{\dagger}$ & \scriptsize[-0.11, +0.07] \\
\addlinespace
\multicolumn{11}{l}{\textit{$g \times a$}}\\
\quad Low capability & -0.23 & \scriptsize[-0.25, -0.20] & +0.03 & \scriptsize[+0.01, +0.05] & +0.47 & \scriptsize[+0.40, +0.53] & -0.09 & \scriptsize[-0.15, -0.02] & +0.31 & \scriptsize[+0.25, +0.37] \\
\quad High capability & +0.13 & \scriptsize[+0.12, +0.13] & -0.03 & \scriptsize[-0.04, -0.03] & -0.09$^{\dagger}$ & \scriptsize[-0.21, +0.03] & -0.04$^{\dagger}$ & \scriptsize[-0.13, +0.05] & -0.18 & \scriptsize[-0.34, -0.02] \\
 
\bottomrule
\end{tabular}}
\end{table}

\subsection{RQ2: joint effects}\label{sec:rq2}

The interaction terms in \cref{tab:regression} show that \AGC{} and \AGT{} are neither simply additive nor mutually compensating. The clearest complementarity concerns recovery. With both active, the low-capability team's FDRT is 7.7\,h, against 12.9\,h with \AGC{} alone and a baseline of 8.9\,h (\cref{tab:levels}), so agentic diagnosis and rollback more than offset the diagnosis penalty of AI-generated failures. For throughput, the interaction is positive in the high-capability team ($\beta_{ga}=+0.13$ for DF): agentic pipeline acceleration and \AGC{}'s extra changes reinforce each other, giving 27.4 deployments per week. In the low-capability team it is negative ($\beta_{ga}=-0.23$), because part of \AGC{}'s DF gain was rework that \AGT{} partly removes.

For change failure rate, H3c predicted that \AGT{} would attenuate the \AGC{} penalty. The model does not support this. In the low-capability team the interaction is positive ($\beta_{ga}=+0.47$): \AGT{} lowers CFR by 29\% without \AGC{} but only by 8\% with it (31.4\% to 28.9\%). In the high-capability team it is not significant. Because the linear interaction term summarizes a nonlinear response, the cell contrast is the more direct evidence. The mechanism is a saturation effect. The probability that a deployment fails, $1-\exp(-q\,\bar e\,n)$ for $n$ changes carrying $\bar e$ escaped defects each, is concave in $n\bar e$. Smaller batches therefore reduce CFR less when \AGC{} has raised $\bar e$.

\subsection{RQ3: the speed--stability trade-off and the role of capability}\label{sec:rq3}

\begin{figure}[!tbp]
\centering
\includegraphics[width=\textwidth]{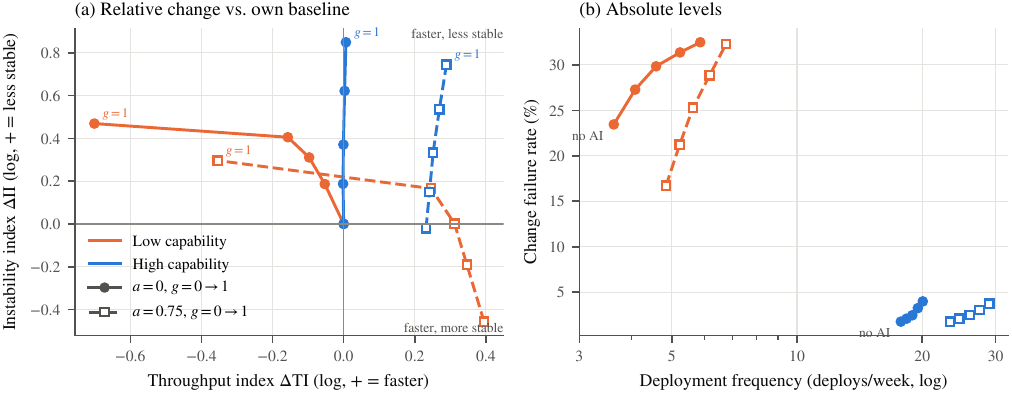}
\caption{Speed--stability trade-off (E1). (a) Throughput and instability indices (\cref{eq:indices}) relative to each team's no-AI baseline; markers trace $g=0,0.25,\dots,1$. (b) Absolute levels of deployment frequency (log scale) and change failure rate for the same scenarios.}
\label{fig:tradeoff}
\end{figure}

\Cref{fig:tradeoff}a places every factorial scenario in the plane spanned by the throughput and instability indices of \cref{eq:indices}, relative to each team's own baseline. \AGC{} alone ($a=0$, solid paths) moves both teams upward: stability deteriorates in both. Neither team gains aggregate throughput, however. The low-capability team loses it steadily ($\Delta\mathrm{TI}=-0.16$ at $g=0.75$) and then sharply once review saturates ($-0.70$ at $g=1$), because planned deployments do not increase while lead time and recovery time lengthen. The high-capability team's index stays at zero: its gains in lead time and planned deployment frequency are offset by longer recovery. \AGT{} ($a=0.75$, dashed paths) shifts both teams to the right, and in the low-capability team also downward. A trade-off in the strict sense, faster but less stable, appears only when both are adopted: at $g=0.75$ for the low-capability team (at $g=1$ its throughput gain disappears as review saturates) and at every $g>0$ for the high-capability team.

\Cref{fig:tradeoff}b shows the same scenarios in absolute levels and qualifies the amplifier proposition (H6). In absolute terms capability strongly dampens AI's harm. At $(g,a)=(0.75,0)$ the CFR increase is 7.9 percentage points in the low-capability team and 1.5 points in the high-capability team. In relative terms the ordering reverses: $+34\%$ against $+84\%$ (\cref{tab:regression}: $\beta_g=+0.38$ vs.\ $+0.83$). Whether AI ``amplifies'' weaknesses or strengths therefore depends on whether effects are measured in levels or in proportions, a choice that current survey evidence leaves implicit.

\Cref{tab:moderators} attributes the difference between the all-low and all-high AI effects to the four capabilities (E3). The capabilities act on different metrics. Platform and observability account for $+4.8$ of the $+4.5$ deployments per week by which the AI throughput gain grows, for most of the reduction in AI-induced CFR ($-4.7$ of $-6.0$ points) and DRR ($-6.5$ of $-9.6$ points), and for the elimination of the recovery-time penalty. They do not, however, reduce failures: their contribution to AI-induced failure intensity is $+0.22$ per week, so their CFR and DRR reductions come from spreading a similar number of failures over more deployments and from rolling back instead of fixing forward. The reductions in failures come from small batches ($-0.33$ per week, together with $-2.0$ and $-3.2$ points of CFR and DRR) and test automation ($-0.32$ per week, about a third of the all-low effect of $+0.90$). Test automation nevertheless slightly \emph{raises} the AI effect on CFR, because with fewer rework deployments the remaining failures are a larger share of all deployments. Review capacity matters almost only for lead time ($-1.5$\,h).

\begin{table}[!tbp]
\centering
\caption{Shapley decomposition of the change in the AI effect ($(g,a)=(0.75,0.5)$ vs.\ $(0,0)$) from the all-low to the all-high capability profile (E3; 100 paired seeds per configuration). Rows between the rules sum to the difference between the last and the first row.}
\label{tab:moderators}
\small
\begin{tabular}{@{}lcccccc@{}}
\toprule
& DF (/week) & CLT (h) & CFR (pp) & FDRT (h) & DRR (pp) & FI (/week) \\
\midrule
AI effect, all moderators low & +2.06 & -0.7 & +7.5 & +0.32 & +10.6 & +0.90 \\
\midrule
\quad Test automation & -0.37 & -0.0 & +0.7 & -0.07 & +0.3 & -0.32 \\
\quad Review capacity & +0.15 & -1.5 & -0.0 & +0.03 & -0.2 & -0.00 \\
\quad Small batches & -0.02 & -0.8 & -2.0 & +0.09 & -3.2 & -0.33 \\
\quad Platform \& observability & +4.76 & +1.9 & -4.7 & -0.45 & -6.5 & +0.22 \\
\midrule
AI effect, all moderators high & +6.59 & -1.1 & +1.5 & -0.08 & +1.0 & +0.47 \\
 
\bottomrule
\end{tabular}
\end{table}

\subsection{RQ4: transmission mechanisms}\label{sec:rq4}

\begin{figure}[!tbp]
\centering
\includegraphics[width=\textwidth]{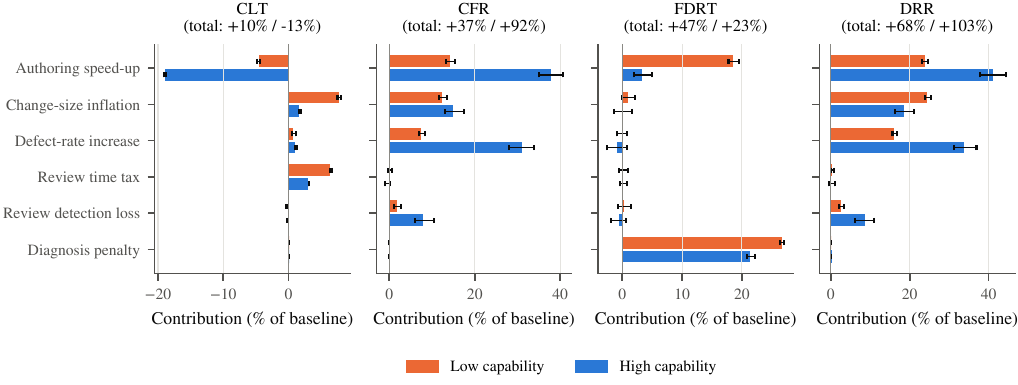}
\caption{Shapley decomposition of the \AGC{} effect ($g=0.75$ vs.\ $0$, $a=0$) into six mechanisms (E2), in percent of each team's baseline. Totals are given as low/high capability. Error bars: 95\% bootstrap CIs over 100 paired seeds.}
\label{fig:shapley}
\end{figure}

\Cref{fig:shapley} decomposes the effect of \AGC{} ($g=0.75$, $a=0$) into its six mechanisms (E2). Three channels stand out.

\emph{Queueing} explains lead time. In the low-capability team, change-size inflation ($+7.8\%$ of baseline CLT) and the review tax ($+6.5\%$) more than offset the authoring speed-up ($-4.6\%$), for a net $+10.2\%$. In the high-capability team, spare review capacity lets the speed-up dominate ($-19.0\%$, net $-13.1\%$).

\emph{Volume and batching} explain much of the instability. The authoring speed-up does not change per-change quality, yet it is the largest single contributor to the CFR increase in both teams ($+14.3\%$ and $+37.9\%$ of baseline) and, together with change-size inflation, to DRR. Under the unbounded-backlog assumption, more changes per week released at an unchanged cadence means more changes per deployment and hence more deployments containing at least one defect (\cref{sec:robustness} examines fixed demand). Change-size inflation adds $+12.6\%$ and $+15.0\%$. The defect-rate multiplier matters most where other defect sources are small: $+31.2\%$ in the high-capability team against $+7.6\%$ in the low-capability team.

\emph{Diagnosis} explains recovery: the diagnosis penalty contributes $+26.7\%$ and $+21.5\%$ of baseline FDRT, and the authoring speed-up adds a further $+18.7\%$ in the low-capability team, because failed deployments then carry more changes to diagnose.

The review tax, the variable closest to ``review latency'' in H7, affects lead time but not stability. The review-detection loss contributes modestly to CFR ($+1.9\%$ and $+8.1\%$).

\subsection{RQ5: dynamics after adoption and their identification}\label{sec:rq5}

\begin{figure}[!tbp]
\centering
\includegraphics[width=\textwidth]{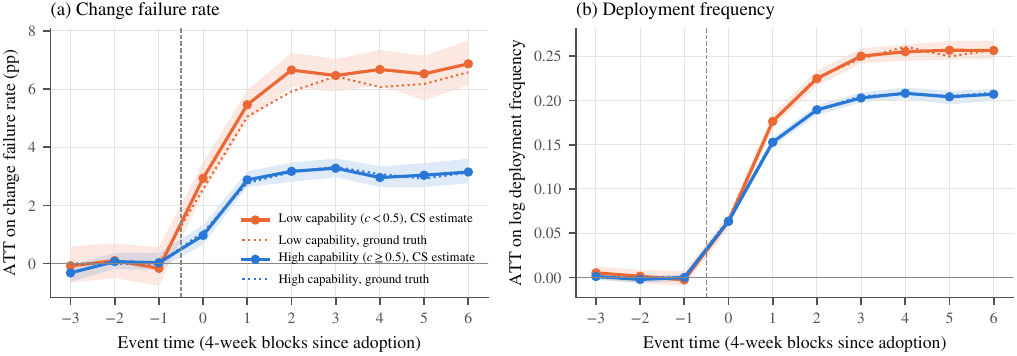}
\caption{Event-study estimates of adoption effects on (a) change failure rate and (b) log deployment frequency by capability subgroup (E5). Solid lines: \citet{callaway2021difference} estimator with never-adopting controls and 95\% team-bootstrap bands (200 replications). Dotted lines: ground truth from counterfactual runs (exactly zero before adoption by construction).}
\label{fig:eventstudy}
\end{figure}

\Cref{fig:eventstudy} shows event-study estimates from the staggered-adoption panel (E5; 5{,}000 teams, 1{,}510 never adopting) together with the ground-truth effects obtained from each team's counterfactual run. Deployment frequency rises monotonically after adoption, with no initial dip, and plateaus after about three blocks. Change failure rate rises with adoption and does not recover within the 28 post-adoption weeks observed. For teams with capability $c\ge0.5$, the true effect peaks at 3.3 points at event time 3 and then lies between 2.9 and 3.1 points. For teams with $c<0.5$ it reaches 6.1--6.6 points and stays there. The J-curve of H8, initial deterioration followed by recovery, is thus at most weakly present: the post-peak decline for high-capability teams (about 0.3 points) is small relative to the confidence bands, and there is no decline for low-capability teams. This persistence follows largely from the assumed dynamics. Adoption is about 95\% complete by event time 3. Learning acts only on the defect, review and diagnosis penalties, not on change volume, size inflation or masking. Its time constant (8--40 weeks) leaves much of the penalty in place within the observation window.

\Cref{tab:estimators} compares the three estimators with the ground truth. The Callaway--Sant'Anna estimator recovers both effects: 4.33 points [4.07, 4.63] against a true 4.14 for CFR, and 0.194 against 0.194 for log DF. Its pre-adoption placebo estimates are within $\pm0.32$ points of zero; one of six, for high-capability teams at $e=-3$ ($-0.32$), has a bootstrap interval that excludes zero. Two-way fixed effects underestimates the CFR effect by 8\% and the DF effect by 16\%, and its confidence intervals exclude the truth; the downward bias is expected when effects grow over event time under staggered adoption \citep{goodmanbacon2021difference,sun2021estimating}. The naive pre/post comparison overstates the CFR effect by 9\% because it attributes the secular rise in defect rates to adoption. The estimator also recovers the heterogeneity by capability: true CFR effects of 5.6 and 2.8 points for low- and high-capability teams are estimated as 6.0 and 2.8.

\begin{table}[!tbp]
\centering
\caption{Recovery of adoption effects (ATT over post-adoption blocks) from the E5 panel of 5{,}000 teams. Brackets: 95\% team-bootstrap CIs.}
\label{tab:estimators}
\small
\begin{tabular}{@{}lcc@{}}
\toprule
Estimator & CFR (percentage points) & log DF \\
\midrule
Ground truth (counterfactual runs) & 4.14 & 0.194 \\
Naive pre/post (adopters only) & 4.53 [4.40, 4.67] & 0.197 [0.194, 0.200] \\
Two-way fixed effects & 3.82 [3.65, 3.95] & 0.164 [0.160, 0.167] \\
\citet{callaway2021difference} & 4.33 [4.07, 4.63] & 0.194 [0.189, 0.198] \\
\bottomrule
\end{tabular}
\end{table}

\subsection{Robustness to parameter uncertainty}\label{sec:robustness}

\Cref{tab:verdicts} summarizes the evidence for each hypothesis. It combines the calibrated effect (E1) with the share of 600 Latin-hypercube parameter draws in which the hypothesized sign holds (E4). Six hypotheses (H1b, H2b, H3a, H4a, H4b, H5) hold in at least 94\% of draws in both profiles, and H2a behaves as its conditional form predicts. H1a holds robustly at low capability but depends on a positive authoring speed-up at high capability. H3b is conditional on capability and the masking rate, H7 is partially supported, and H3c, H6 and H8 are not supported as stated.

\begin{table}[!tbp]
\centering
\caption{Hypothesis verdicts. ``Calibrated effect'' is the E1 change at 0.75 adoption (low / high capability). ``Share of draws'' is the percentage of 600 E4 parameter draws in which the hypothesized sign holds (low / high capability).}
\label{tab:verdicts}
\small
\begin{tabularx}{\textwidth}{@{}lp{3.6cm}cL@{}}
\toprule
ID & Calibrated effect & Share of draws (\%) & Verdict \\
\midrule
H1a & DF $+44\%$ / $+10\%$ & 97.8 / 79.2 & Supported at low capability, where the gain is entirely rework (planned DF unchanged); at high capability conditional on $\lambda_S>0$ and on AI raising change volume (E6). \\
H1b & DF $+34\%$ / $+31\%$ & 97.7 / 100 & Supported. \\
H2a & CLT $+10\%$ / $-13\%$ & 13.3 / 57.3 {\scriptsize(CLT $\downarrow$)} & Supported as conditional: net effect set by review capacity and speed-up; saturation at high $g$; the lead-time penalty requires AI to raise change volume (E6). \\
H2b & CLT $-22\%$ / $-8\%$ & 100 / 100 & Supported. \\
H3a & CFR $+7.9$ / $+1.5$\,pp & 97.5 / 97.3 & Supported. \\
H3b & CFR $-6.7$ / $-0.1$\,pp & 99.0 / 49.7 & Conditional: holds at low capability (partly a denominator effect); at high capability the sign is set by the masking rate $\mu$. \\
H3c & $\beta_{ga}=+0.47$ / $-0.09^\dagger$ & 5.0 / 37.8 & Not supported (saturation of per-deployment failure probability). \\
H4a & FDRT $-38\%$ / $-29\%$ & 100 / 100 & Supported. \\
H4b & FDRT $+45\%$ / $+24\%$ & 98.8 / 94.0 & Supported. \\
H5 & DRR $+21.1$ / $+1.2$\,pp & 97.8 / 96.5 & Supported. \\
H6 & see \cref{tab:moderators} & -- & Not supported as stated: high-capability teams also lose stability. Capability dampens harm in absolute but not relative terms, and raises the absolute throughput gain. \\
H7 & see \cref{fig:shapley} & -- & Partially: size mediates CLT and CFR; review latency mediates CLT only; unhypothesised volume/batching channel. \\
H8 & see \cref{fig:eventstudy} & -- & Not supported within 28 weeks under the assumed learning dynamics: CFR effects persist; weak post-peak decline at high capability. \\
\bottomrule
\end{tabularx}
\end{table}

\begin{figure}[!tbp]
\centering
\includegraphics[width=\textwidth]{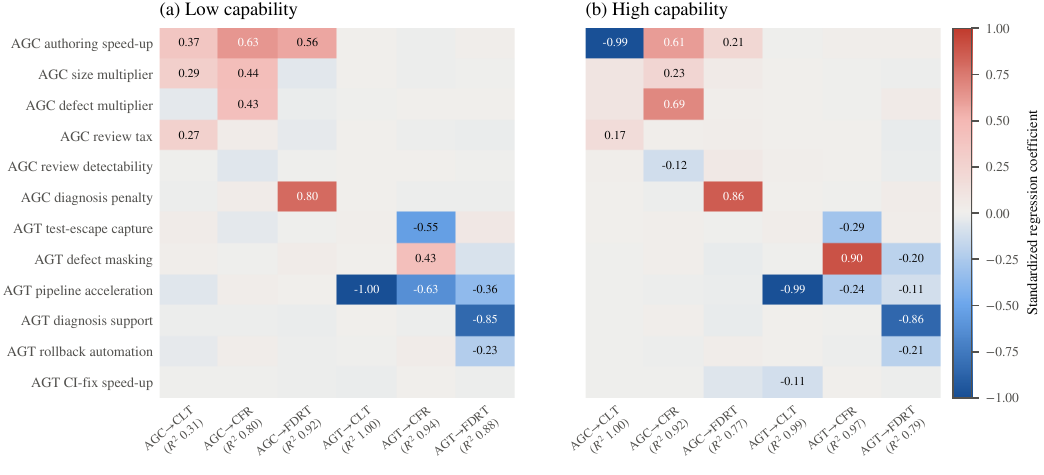}
\caption{Standardized regression coefficients of \AGC{} and \AGT{} effects on the twelve AI mechanism parameters (E4; 600 Latin-hypercube draws $\times$ 6 replications). Coefficients with $|\text{SRC}|\ge0.1$ are labeled; $R^2$ of each linear meta-model is given below the effect.}
\label{fig:gsa}
\end{figure}

\Cref{fig:gsa} identifies the parameters that drive each effect. \AGT{}'s effect on CFR in the high-capability team is governed almost entirely by the masking probability $\mu$ (SRC $=0.90$, $R^2=0.97$). \AGT{} lowers CFR in 100\% of draws with $\mu<0.05$, in 72\% with $0.10\le\mu<0.15$, and in 2.5\% with $\mu\ge0.20$. Its effect on failure intensity turns adverse at an even lower masking rate: \AGT{} lowers failure intensity in only 13\% of draws with $0.10\le\mu<0.15$. \AGC{}'s effect on lead time in the high-capability team is governed by the authoring speed-up (SRC $=-0.99$); a lead-time reduction occurs in every draw with $\lambda_S>0.3$ and in none with $\lambda_S<0$. In the low-capability team the same effect is poorly described by a linear meta-model ($R^2=0.31$) because of review saturation. There, total deployment frequency rises in 93\% of the draws in which \AGC{} \emph{slows} authoring ($\lambda_S<0$, consistent with \citealp{becker2025measuring}), because defects generate rework deployments.

\paragraph{Fixed demand (E6).} The unbounded-backlog assumption lets faster authoring raise change volume. When AI instead only shortens the authoring of a given number of changes (\cref{tab:fixeddemand}), the stability effects of \AGC{} keep their sign but shrink. CFR rises by 6.1 points ($+26\%$) in the low-capability team and by 0.7 points ($+41\%$) in the high-capability team, against 7.9 and 1.5 points under the unbounded backlog. Failure intensity rises by 57\% and 42\%, rework rate by 44\% and 50\%, and recovery time by 26\% and 20\%. The volume-dependent findings change. The low-capability team's lead time no longer increases ($-2.5\%$), because the review queue carries no extra load, and the high-capability team's deployment frequency no longer rises. The low-capability team's DF still rises by 24\%, entirely through rework. \AGT{}-only results are unaffected, because they do not involve AI-generated changes. At the calibrated values, H3a, H4b and H5 therefore do not depend on the volume assumption, whereas H1a at high capability and the lead-time penalty in H2a do.

\subsection{Implications for empirical study design}\label{sec:results-design}

\begin{figure}[!tbp]
\centering
\includegraphics[width=0.6\textwidth]{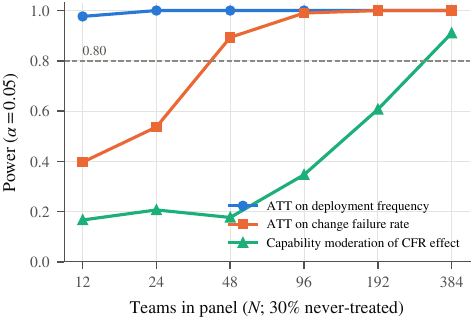}
\caption{Power of the \citet{callaway2021difference} estimator to detect adoption effects, and their moderation by capability ($c\ge0.5$ vs.\ $c<0.5$), as a function of panel size (E5; 300 panels per size, 199 bootstrap replications each, $\alpha=0.05$).}
\label{fig:power}
\end{figure}

\Cref{fig:power} reports the power of the Callaway--Sant'Anna estimator in panels of $N$ teams observed for 52 weeks with the same adoption structure as E5. These figures assume what the simulation provides: known adoption dates, 30\% never-adopting teams, thirteen 4-week blocks, error-free measurement, and effects of the size the model generates. Under these conditions an adoption effect on deployment frequency is detected with power $\ge0.98$ even with 12 teams. The effect on change failure rate, about 4 percentage points, needs about 48 teams for power 0.89. Testing whether capability moderates the CFR effect, the amplifier hypothesis at team level, requires far more: power is 0.61 with 192 teams and 0.91 with 384. Estimates of moderation power are unstable at or below 48 teams, where capability subgroups are very small. Coverage of the ground truth by the bootstrap intervals lies between 93\% and 96\% for the CFR effect and between 94\% and 100\% for the DF effect. Measurement error in AI attribution or failure linkage would raise the required sample sizes. Single-team before/after studies, the dominant design in the current literature (\cref{tab:evidence}), cannot resolve stability effects of this size.

\section{Discussion}\label{sec:discussion}

\subsection{Principal findings and why they arise}\label{sec:principal}

The results come down to three structural properties of delivery systems. The directions of the effects follow from these properties rather than from particular parameter values; their magnitudes, and the exceptions listed in \cref{tab:verdicts}, do depend on parameters.

\paragraph{AI speed is absorbed by the binding constraint.} Faster authoring raises the arrival rate at the next stage. Whether that shortens lead time depends on the remaining capacity of review and deployment, and waiting time grows nonlinearly as utilization approaches one \citep{little1961proof,kingman1961single}. The low-capability team converts AI-generated volume into review queuing and, once review saturates, into lead times at least five times the baseline. The high-capability team, with spare review capacity and frequent deployments, converts the same volume into shorter lead times. AI-generated code therefore accelerates delivery only where the constraint lies elsewhere, the familiar lesson of flow-based product development \citep{reinertsen2009principles} applied to a new source of work.

\paragraph{DORA ratios mix volume, batching and quality.} Change failure rate and rework rate are per-deployment ratios, and deployment frequency counts rework, so all three respond to changes in the number and size of deployments as well as to quality. Several results in \cref{sec:results} are denominator effects of this kind: \AGT{}'s CFR reduction in the low-capability team, the role of the authoring speed-up in the CFR increase, the CFR effects of test automation and platform capability, and the rework-driven DF gain of \AGC{}. A team or vendor that reports rising deployment frequency and falling change failure rate after AI adoption may therefore be describing more rework and smaller batches rather than better software. Planned deployment frequency and failure intensity separate these cases.

\paragraph{Autonomy moves risk from speed to oversight.} Agentic AI improved throughput and recovery in every E1 configuration and in at least 97\% of parameter draws, consistent with its role as automation of hand-offs. Its stability effect depends on how often autonomous remediation hides defects instead of fixing them. Because a stronger test suite detects more defects, it also gives an agent more opportunities to mask one, so mature teams are the most exposed. This result was not anticipated in our hypotheses and is the most consequential for practice.

\subsection{Comparison with prior evidence}\label{sec:comparison}

\emph{DORA 2024 and 2025.} The two DORA reports appear to contradict each other: throughput fell with AI adoption in 2024 and rose in 2025, while stability fell in both \citep{dora2024accelerate,dora2025aiassisted}. The model generates both patterns from the same mechanisms. \AGC{} without agentic support, in teams near review capacity, lowers aggregate throughput and stability (\cref{fig:tradeoff}a, solid orange path), as in 2024. \AGC{} combined with agentic automation raises throughput and lowers stability in both capability profiles, as in 2025. One possible explanation of the reversal, which the model suggests but cannot test, is therefore a shift in the adoption mix between the survey waves from code assistants towards agentic tooling. The model also gives mechanism-level support to DORA's 2024 conjecture that AI enlarges batches: size inflation and more changes per deployment carry a large share of the instability effect (\cref{fig:shapley}).

\emph{Controlled field evidence.} \Citet{gunawan2026assessing} report lead time $-39.2\%$, deployment frequency $+61.9\%$, CFR 14.3\% to 8.7\% and recovery time $-33\%$ for an AI-augmented DevSecOps pipeline, an intervention closer to our \AGT{} than to \AGC{}. The model's \AGT{}-only results for the low-capability team point in the same direction with comparable magnitudes: CLT $-22\%$, DF $+34\%$, CFR $-29\%$ and FDRT $-38\%$. Our decomposition adds two caveats: part of such a CFR reduction can be a denominator effect, and it would reverse under high masking rates, which a two-sprint study cannot observe.

\emph{Productivity and mining studies.} The perception gap documented by \citet{becker2025measuring}, where developers believed they were faster while being slower, has a delivery-level analogue in the model: deployment frequency can rise even when AI slows authoring, because of rework. The first-run CI failure rate of all changes rises from 17.8\% to 26.1\% at $g=0.75$ in the low-capability team, in line with the importance of CI failures for agentic pull requests \citep{ehsani2026where}. The model makes explicit how the survival of AI-introduced issues \citep{liu2026debt} and the extra revisions agentic pull requests need \citep{watanabe2026agentic} would appear in the operational metrics that organizations track: rework rate, failure intensity and recovery time.

\subsection{Unexpected findings}\label{sec:unexpected}

Four results were not anticipated by the hypotheses.
\begin{enumerate}[label=(\arabic*)]
\item Agentic AI can \emph{increase} escaped defects in high-capability teams, through masking (\cref{sec:robustness}).
\item Agentic validation does not offset the CFR penalty of AI-generated code (H3c). The reason is the concavity of per-deployment failure probability, not a weak agent: once AI-generated code has raised the defect content of deployments, splitting them into smaller ones helps less.
\item Capability reduces AI's stability cost in levels but not in proportions, so the answer to ``does AI help strong teams more?'' depends on the scale of measurement. The same holds for throughput: capability raises the absolute DF gain of AI but not its relative gain.
\item No J-curve recovery of stability appears within 28 weeks. This is largely a consequence of the assumed learning dynamics (\cref{sec:rq5}), but it cautions against assuming that the stability cost of adoption is transitional.
\end{enumerate}

\subsection{Implications}\label{sec:implications}

\paragraph{Theory.} The study supports treating AI adoption in software delivery as two constructs with different causal footprints: an artefact effect (\AGC{}) acting on volume, size and defect content, and an actor effect (\AGT{}) acting on hand-offs, verification and recovery. It reformulates the amplifier proposition as a testable, scale-dependent moderation hypothesis: capability reduces the absolute stability cost of AI and increases its absolute throughput gain, while leaving the proportional stability cost at least as large. It also treats DORA metrics as outputs of a queuing and batching system, so predictions about AI must be made jointly for arrival rates, capacities and cadences.

\paragraph{Measurement and method.} Studies of AI adoption should report planned deployment frequency and failure intensity alongside the standard metrics, decompose lead time into coding, review and release stages, and measure \AGC{} and \AGT{} separately at change and action level (\cref{app:protocol}). In the simulated panel, heterogeneity-robust staggered difference-in-differences recovered effects that two-way fixed effects underestimated by 8--16\% and naive pre/post comparisons overstated. Under the idealized conditions of \cref{sec:results-design}, detecting a stability effect of about four points required about 50 teams and testing its moderation by capability several hundred. Real measurement error would raise both numbers. This argues for multi-organization consortia or telemetry-sharing arrangements rather than single-company case studies.

\paragraph{Practice and management.} The results suggest three priorities before AI-generated code is scaled. First, reduce the number of failures AI adds: keep changes small, including AI-generated ones, and strengthen automated tests. In the moderator decomposition these were the only capabilities that reduced AI-induced failure intensity. Second, make failures cheap: automated deployment, fast detection and reliable rollback eliminated the recovery-time penalty and lowered AI-induced CFR and rework, although mainly by spreading failures over more deployments. Third, size review capacity to keep utilization well below saturation after the expected increase in change volume. Agentic remediation needs explicit guardrails against masking: agents should not modify or disable tests without human approval, agent-authored fixes should be tracked separately, and escaped defects should be monitored by remediation source. Managers should not read rising deployment frequency or falling change failure rate as evidence of AI value unless planned deployments and failure intensity move in the same direction.

\paragraph{Policy and governance.} For public-sector and regulated organizations that procure AI development tools at scale, the results argue for outcome-based evaluation over activity metrics. Adoption programs should require the telemetry needed to separate planned from unplanned deployments and to attribute changes and actions to AI, and should be evaluated with staggered rollouts that provide credible comparison groups. Governance frameworks for agentic AI \citep{adabara2025trustworthy} should treat autonomous remediation in CI/CD as a controlled capability, with audit trails and permission scopes.

\paragraph{Research agenda.} The model makes its uncertain inputs explicit. The most decision-relevant are the masking rate of agentic remediation ($\mu$), the defect-rate and size multipliers of AI-generated code ($\lambda_D,\lambda_L$), the review tax ($\lambda_R$), and whether AI raises change volume or only shortens authoring. Each can be estimated from repository, CI and planning data, which would turn the conditional verdicts of \cref{tab:verdicts} into firm predictions. The protocol in \cref{app:protocol} and the power analysis in \cref{sec:results-design} provide the blueprint for a field study, extending the multi-site validation protocol proposed in \citet{alenezi2026auditable}.

\section{Threats to Validity}\label{sec:threats}

\paragraph{Construct validity.}
The model deliberately abstracts a software delivery system into a single team handling independent changes through one review stage, one test suite, and a deployment train. It therefore omits several features of production environments, including cross-service dependencies, feature flags, canary releases, reviewer learning, and interactions among concurrent incidents. Some simplifications also affect how individual DORA outcomes are represented. Rework deployments contain no regular changes and cannot themselves fail, while rolled-back changes are not subsequently redeployed; consequently, a higher rollback probability can mechanically reduce measured rework. In addition, a deployment is classified as failed only when a defect manifests immediately, whereas the empirical protocol in \cref{app:protocol} attributes failures occurring within seven days to the originating deployment. Some defects treated as rework in the simulation could therefore be classified as change failures under the empirical protocol. The baseline model also assumes an unbounded backlog, so reductions in authoring time can translate directly into greater change volume; E6 explicitly relaxes this assumption. Finally, the capability profiles represent stylized organizational endpoints rather than empirically estimated clusters. We mitigate these limitations by grounding each mechanism in a published argument or empirical finding, keeping the model sufficiently parsimonious for systematic verification (\cref{app:verification}), and interpreting the results primarily in terms of causal direction, interaction, and mechanism rather than exact effect magnitude.

\paragraph{Internal validity.}
The principal internal-validity threat is uncertainty in the mechanism parameters. None of the AI-specific parameters has been directly estimated for the delivery process represented here, and some, particularly the masking rate, remain assumption-driven. We therefore use deliberately broad parameter ranges and evaluate not only effect magnitude, but also sign robustness and the parameters most responsible for variation in each outcome. This approach addresses parametric uncertainty but not structural uncertainty: sensitivity analysis cannot determine whether an omitted or incorrectly specified mechanism would materially change the conclusions. The reported verdicts should therefore be read as conditional on the model structure. In particular, the robustness shares in \cref{tab:verdicts} describe the proportion of sampled parameter configurations producing a given directional result; they are not posterior probabilities that the corresponding hypotheses are true. The learning dynamics introduced in E5 are similarly assumed rather than empirically estimated.

\paragraph{External validity.}
The model is not calibrated to a specific organization, repository, or delivery architecture, and its results should not be interpreted as universal effect estimates. Outcomes may differ substantially in settings with strong regulatory approval gates, large monorepos, tightly coupled service architectures, or safety-critical release processes. The staggered-adoption analysis also assumes that adoption timing is independent of otherwise unobserved team characteristics. In practice, more capable or better-resourced teams may adopt AI earlier, creating selection effects that can be mistaken for treatment effects. This possibility strengthens the importance of credible comparison groups, pre-trend diagnostics, and designs that distinguish adoption effects from pre-existing organizational capability.

\paragraph{Conclusion validity.}
Simulation uncertainty is modest relative to the calibrated treatment effects because the factorial experiments use 100 replications per cell, paired common random numbers, and bootstrap uncertainty intervals. The global sensitivity analysis is necessarily noisier because each sampled parameter configuration is evaluated with only six replications. Moreover, the study estimates multiple outcomes, mechanisms, interactions, and capability-dependent effects. We therefore avoid interpreting isolated significance tests as decisive evidence and instead emphasize patterns that persist across DORA metrics, organizational capability profiles, experimental designs, and parameter draws.

\paragraph{Evidence synthesis.}
The evidence review was structured to support traceability and model construction, but it was not conducted as a formal systematic review. Requiring consequential claims to be traceable to primary or authoritative sources improves evidential reliability but may exclude relevant findings reported only in secondary literature, industry case studies, or emerging preprints. The evidence base is also evolving rapidly as stronger field studies of AI-assisted and agentic software engineering become available. Accordingly, both the parameterization and the set of mechanisms represented in the model should be revisited as new empirical evidence accumulates.

\section{Conclusion}\label{sec:conclusion}

Asking whether AI improves DevOps performance treats two different interventions as one. By separating AI-generated code from agentic AI and modelling how each flows through a delivery pipeline, this study shows that the two have different causal footprints and that DORA metrics respond to them through queuing, batching and oversight as much as through code quality. Within the model, AI-generated code raises instability under both capability profiles and, when it raises change volume, shortens lead time only where review capacity can absorb it. Agentic AI improves throughput and recovery but can raise escaped defects where autonomous remediation masks failures. Engineering capability reduces the absolute cost of AI adoption without removing its proportional stability cost. These conclusions follow from the mechanisms that current evidence proposes and hold across most of the parameter space examined, with the exceptions identified in \cref{tab:verdicts}. They remain model-derived and need field testing. The paper provides the operational definitions, estimator and sample-size requirements for that test. The most informative next step is multi-organization telemetry studies that directly measure the masking rate of agentic remediation and the defect and size profile of AI-generated changes.

\bibliographystyle{plainnat}
\bibliography{references}

\appendix
\section{Measurement Protocol for the Empirical Follow-up}\label{app:protocol}

The simulation identifies which quantities an empirical study must measure to separate the effects studied here. \Cref{tab:protocol} gives operational definitions at the level of changes, deployments and incidents. Three rules follow from the results. First, \AGC{} and \AGT{} are measured separately and at the lowest available unit. Second, every ratio metric is reported with its numerator and denominator, so that denominator effects can be detected. Third, the unit of analysis is the team-period, with individual developers never evaluated, to avoid gaming and to respect the team-level scope of DORA metrics \citep{forsgren2018accelerate}.

\begin{table}[!htbp]
\centering
\caption{Operational definitions for field measurement.}
\label{tab:protocol}
\small
\begin{tabularx}{\textwidth}{@{}p{3.2cm}Lp{3.1cm}@{}}
\toprule
Construct & Operationalisation & Source \\
\midrule
\AGC{} share $g$ & Share of merged changes (or changed lines) attributed to AI. Attribution rules in priority order: (1) agent-authored PR (bot account or agent co-author trailer); (2) assistant telemetry of accepted suggestions mapped to commits; (3) author self-label in the PR template. Report the rule used, and test thresholds of 25/50/75\% of lines. & VCS and PR metadata, assistant telemetry \\
\AGT{} intensity $a$ & Agent-executed actions per 100 deployments, by stage: test generation, CI repair, pipeline actions, incident triage/remediation/rollback; with human-approval flag and outcome (success, reverted, failed). & Agent logs, CI/CD audit logs, incident tooling \\
Adoption time $t_0$ & First week in which $g$ or $a$ exceeds a pre-registered threshold (e.g., 10\% of PRs). & Telemetry \\
DF, planned DF & Production deployments per week; planned DF excludes deployments linked to an incident or labelled hotfix. & Deployment log \\
CLT & $t^{\text{dep}}-t^{\text{first commit}}$ per change; median and 75th percentile; decomposed into coding, review pickup, review, merge-to-deploy. & VCS, PR, deployment log \\
CFR, FI & Deployments linked to a production failure requiring remediation within 7 days (sensitivity: 3 and 14 days), divided by deployments (CFR) and by weeks (FI). & Incident--deployment links \\
FDRT & Time from failure detection to restoration, for change-caused failures only; median. & Incident tooling \\
DRR & Unplanned deployments resulting from a production incident, divided by deployments. & Deployment log, incident links \\
Mechanism variables & Change size; review waiting and service time; first-run CI failure rate; escaped defects (post-merge bug links per change); changes per deployment; human intervention rate on agent-authored changes. & VCS, CI, issue tracker \\
Moderators & Test automation (coverage, gating), review capacity (reviewer-hours per PR), change-size distribution, deployment interval, detection time, rollback availability. & CI configuration, telemetry, short survey \\
\bottomrule
\end{tabularx}
\end{table}

\paragraph{Mapping to the simulation.} The simulation implements a subset of these definitions. It counts a deployment as failed only when a defect manifests immediately, and counts latent defects that surface later as rework, whereas the seven-day linkage window above would classify some of them as change failures. Field CFR measured with a seven-day window should therefore be expected to exceed the model's CFR, and some hotfix deployments would then count both as rework and as the remediation of a failed deployment.

\section{Model Verification}\label{app:verification}

\paragraph{Invariance.} Across 20 seeds and both capability profiles, the five DORA metrics with $g=0$ were bitwise identical when the six \AGC{} parameters were set to values far from their baselines ($\lambda_S=0.1$, $\lambda_L=1.9$, $\lambda_D=1.7$, $\lambda_R=1.5$, $\lambda_P=0.6$, $\lambda_G=0.8$). With $a=0$ they were likewise identical when the six \AGT{} parameters were changed ($\eta_T=0.7$, $\mu=0.3$, $\eta_P=0.2$, $\eta_O=0.6$, $\eta_{RB}=0.8$, $\eta_F=0.5$). With every \AGC{} mechanism disabled, runs with $g=0.75$ were identical to runs with $g=0$ (200 paired seeds per profile), which confirms that \AGC{} affects outcomes only through the specified mechanisms. In the E5 panel, the pre-adoption weekly deployments, failures and rework deployments of all 3{,}490\ adopting teams are bitwise identical to those of their counterfactual runs, so the ground-truth effects in \cref{sec:rq5} contain no pre-adoption noise.

\paragraph{Queueing.} \Cref{tab:verification} compares simulated mean review waiting times for the low-capability team ($n_r=2$ reviewers) with the Allen--Cunneen approximation
\[
W_q\approx \frac{C(n_r,u)\,\E[S]}{n_r-u}\cdot\frac{c_a^2+c_s^2}{2},\qquad u=\nu\,\E[S],
\]
where $\nu$ is the arrival rate of changes at review, $\E[S]$ the mean review time, $u$ the offered load, $C(n_r,u)$ the Erlang-C probability of waiting, and $c_a^2$, $c_s^2$ the squared coefficients of variation of inter-arrival and review times \citep{allen1990probability}. Utilisation is $\rho=u/n_r$. The approximation reproduces the steep, nonlinear growth of waiting time with utilisation that drives the lead-time results. Simulated waits lie 16--33\% below the approximation. The approximation treats arrivals as a renewal process, whereas arrivals from a few developers working sequentially are correlated, and it is itself approximate for small numbers of servers; the agreement in growth pattern is what matters for the results.

\begin{table}[!htbp]
\centering
\caption{Review-queue verification (low-capability team, $a=0$, 40 seeds per row).}
\label{tab:verification}
\small
\begin{tabular}{@{}ccccccc@{}}
\toprule
\AGC{} share $g$ & Utilisation $\rho$ & $c_a^2$ & $c_s^2$ & $W_q$ simulated (h) & $W_q$ Allen--Cunneen (h) & Difference (\%) \\
\midrule
0.00 & 0.37 & 0.89 & 1.11 & 0.11 & 0.17 & -33 \\
0.25 & 0.48 & 0.93 & 1.23 & 0.31 & 0.42 & -25 \\
0.50 & 0.62 & 0.92 & 1.24 & 0.79 & 1.01 & -22 \\
0.75 & 0.81 & 0.88 & 1.18 & 2.72 & 3.23 & -16 \\
0.90 & 0.94 & 0.91 & 1.13 & 10.74 & 13.91 & -23 \\
 
\bottomrule
\end{tabular}
\end{table}

\section{Fixed-Demand Variant}\label{app:fixeddemand}

\Cref{tab:fixeddemand} reports the E1 anchor cells re-run under fixed demand (E6). AI shortens the authoring of each change, but developers start new changes at the human pace, so change volume is unchanged. Compare with \cref{tab:levels}.

\begin{table}[!htbp]
\centering
\caption{Anchor scenarios under fixed demand (E6; 100 runs per cell; means). Changes: changes authored per week.}
\label{tab:fixeddemand}
\resizebox{\textwidth}{!}{%
\begin{tabular}{@{}lcccccccc@{}}
\toprule
Scenario & Changes (/wk) & DF (/wk) & Planned DF (/wk) & CLT (h) & CFR (\%) & FDRT (h) & DRR (\%) & FI (/wk) \\
\midrule
\multicolumn{9}{@{}l}{\textit{Low capability team}}\\
\quad Baseline & 26.8 & 3.6 & 2.5 & 15.7 & 23.5 & 8.9 & 31.0 & 0.85 \\
\quad AGC only & 26.6 & 4.5 & 2.5 & 15.3 & 29.6 & 11.3 & 44.5 & 1.34 \\
\quad AGT only & 26.8 & 4.8 & 4.0 & 12.3 & 16.7 & 5.6 & 17.4 & 0.81 \\
\quad AGC + AGT & 26.7 & 5.6 & 4.0 & 11.9 & 24.4 & 6.8 & 28.1 & 1.36 \\
\addlinespace
\multicolumn{9}{@{}l}{\textit{High capability team}}\\
\quad Baseline & 38.1 & 17.8 & 17.5 & 5.6 & 1.8 & 2.3 & 1.3 & 0.31 \\
\quad AGC only & 38.2 & 17.9 & 17.5 & 4.9 & 2.5 & 2.7 & 2.0 & 0.44 \\
\quad AGT only & 38.2 & 23.3 & 23.0 & 5.2 & 1.7 & 1.6 & 1.3 & 0.41 \\
\quad AGC + AGT & 38.2 & 23.4 & 23.0 & 4.4 & 2.6 & 1.9 & 1.8 & 0.60 \\
 
\bottomrule
\end{tabular}}
\end{table}

\end{document}